\documentclass{aa}
\usepackage{graphicx}
\begin{document}
\title{Galaxy mergers with various mass ratios:\\ properties of remnants}
\author{F. Bournaud \inst{1}, C. J. Jog \inst{2}, and F. Combes \inst{1}}
\offprints{F. Bournaud \email{Frederic.Bournaud@obspm.fr}}
\institute{Observatoire de Paris, LERMA, 61 Av. de l'Observatoire, F-75014, Paris, France
\and
Department of Physics, Indian Institute of Science, Bangalore 560012, India}

\date{Received; accepted}

\abstract{We study galaxy mergers with various mass ratios using N-body simulations, with an emphasis on the unequal-mass mergers in the relatively unexplored range of mass-ratios 4:1--10:1. Our recent work (Bournaud et al. 2004) shows that the above range of mass ratio results in hybrid systems with spiral-like luminosity profiles but with elliptical-like kinematics, as observed in the data analysis for a sample of mergers by Jog \& Chitre (2002). In this paper, we study the merger remnants for mass ratios from 1:1 to 10:1 while systematically covering the parameter space. We obtain the morphological and kinematical properties of the remnants, and also discuss the robustness and the visibility of disks in the merger remnants with a random line-of-sight. We show that the mass ratios 1:1--3:1 give rise to elliptical remnants whereas the mass ratios 4.5:1--10:1 produce hybrid systems with mixed properties. We find that the transition between disk-like and elliptical remnants occurs between a narrow mass range of 4.5:1--3:1. The unequal-mass mergers are more likely to occur than the standard equal-mass mergers studied in the literature so far, and we discuss their implications for the evolution of galaxies.
}
\authorrunning{Bournaud et al.}
\titlerunning{}
\maketitle


\section{Introduction}

Mergers between galaxies are known to be frequent and can lead to a significant dynamical and morphological evolution of galaxies. Numerical simulations of mergers of two equal-mass spiral galaxies have been studied extensively (e.g., Barnes \& Hernquist 1991, Barnes 1992). These have been shown to give rise to pressure-supported remnants with an $r^{1/4}$ radial mass profile, as observed in elliptical galaxies (de Vaucouleurs 1977). These so-called major mergers result in a dramatic violent relaxation leading to the formation of an elliptical galaxy, as was proposed theoretically (Toomre 1977). Recently, mergers of galaxies with comparable masses with the mass ratios in the range 1:1--3:1 or 1:1--4:1 have also been studied by N-body simulations (Bendo \& Barnes 2000, Cretton et al. 2001, Naab \& Burkert 2003). These also mostly result in elliptical-like remnants, but which can be disky or boxy.

These models were largely motivated by the observations of infrared-bright, ultra-luminous galaxies, which appear to be the result of comparable-mass galaxy mergers. A few of these mergers show an $r^{1/4}$ de Vaucouleurs profile typical of elliptical galaxies (e.g., Schweizer 1982, Stanford \& Bushouse 1991, Chitre \& Jog 2002). Thus, the main aim of these theoretical studies seems to be to show that merger remnants with elliptical-like mass profiles can form.

At the other extreme end of the range of mass ratios, the so-called minor mergers between a large galaxy and a satellite galaxy with a ratio of 10:1 or more have also been studied numerically (Quinn, Hernquist \& Fullagar 1993, Walker, Mihos \& Hernquist 1996, Velaquez \& White 1999). These result in hot, thickened disk galaxies which still have an exponential mass distribution, as in an isolated spiral galaxy (Freeman 1970).

Surprisingly, the large intermediate range of mass ratios (4:1--10:1) has not been explored in the literature, perhaps because there was no clear observational motivation for doing so. However, given that the observed mass spectrum of galaxies peaks at lower masses (i.e., the Schechter luminosity function, see e.g., Binney \& Tremaine 1987), it is obvious that mergers with this mass range are more likely to occur than the equal-mass cases that have been studied commonly in the literature so far. Hence, such unequal-mass mergers need to be studied in detail. Note that these must be even more important in the early evolution of galaxies.

This new mass range (4:1--10:1) was explored recently in numerical simulations by Bournaud, Combes \& Jog (2004) who showed that the above range of mass ratios can result in ''hybrid'' systems with spiral-like morphology but elliptical-like kinematics. These results explain well the observed properties of a sample of advanced mergers analyzed by Jog \& Chitre (2002), and the simulations by Bournaud et al. (2004) were motivated by these observations.

In this paper, we study galaxy mergers with various mass ratios, mainly focus on unequal-mass mergers in this new range of mass ratios, and systematically cover the detailed parameter space-such as the orbital parameters, study the morphology and the global kinematics of the remnants. We show that there is a well-defined small mass range, corresponding to a ratio of 3:1--4.5:1 for the stellar masses, over which the remnants show a transition from a disk-like to an elliptical morphology. We also study additional properties like the disk visibility, diskiness/boxiness of the thick disk and bulge, and the gas response. Further, we study the implications of these for galaxy evolution, including the formation of S0s, and also discuss how multiple unequal-mass mergers could be the progenitors of elliptical galaxies.

Sect.~2 contains the details of N-body simulations. In Sect.~3 we analyze the properties of the merger remnants as a function of the mass ratios. In Sect.~4, we study in more detail the properties of the merger remnants in the new range of mass ratios 4:1--10:1. Their implications for galaxy evolution are discussed in Sect.~5. Sect.~6 contains a brief summary of results from this paper.


\section{N-body simulations of galaxy mergers}

\subsection{Code description}

We have used the N-body FFT code of Bournaud \& Combes (2003). The gravitational fields are computed on a grid of size $256^3$, with a resolution of 700pc. We used $10^6$ particles for the most massive galaxy. The number of particles used for the other galaxy is proportional to its mass. Star formation and time-dependent stellar mass-loss schemes used are as described in Bournaud \& Combes (2002). The star formation rate is computed according to the generalized Schmidt law (Schmidt 1959): the local star formation rate is assumed to be proportional to ${\mu_g}^{b}$, where $\mu_g$ the is local two-dimensional density of gas. We chose $b=1.4$, as suggested by the observational results of Kennicutt (1998).
 The dissipative dynamics of the ISM has been accounted for by the sticky-particles scheme described in the  Appendix~A of Bournaud \& Combes (2002). In this paper we employ elasticity parameters $\beta_t$=$\beta_r$=0.8. 

\subsection{Physical model for colliding galaxies}
Each galaxy is initially made-up of a stellar and gaseous disk, a spherical bulge, and a spherical dark halo. The visible mass of the main galaxy is $2\times 10^{11}$ M$_{\sun}$. Its stellar disk is a Toomre (1964) disk of radial scalelength 5kpc, truncated at 15 kpc. Gas represents 8\% of the disk mass, and is distributed in a disk of 30 kpc radius. The bulge and dark halos are Plummer spheres of radial scalelengths 3 kpc and 40~kpc respectively. The bulge-to-total mass ratio is 0.17 (bulge-to-disk: 0.2), and the dark-to-visible mass ratio inside the stellar disk radius is 0.5. The initial velocities of particles are computed as in Bournaud \& Combes (2003). The initial value of the Toomre parameter is Q=1.7 over the whole disk.

The radial distribution of matter in the other galaxy has been scaled by the square root of its stellar mass. Its gas and dark matter content have been varied according to Table~\ref{params}.

\begin{table*}
\centering
\begin{tabular}{lcc}
\hline
\hline
Galaxy          & Gas mass fraction & Dark-to-visible ratio \\
                & in the disk       & inside the stellar radius \\
\hline
Main galaxy     &   8\%             & 0.5 \\
Companion 1:1   &   8\%             & 0.5 \\
Companion 2:1   &   8\%             & 0.5 \\
Companion 3:1   &  10\%             & 0.6 \\
Companion 4.5:1 &  11\%             & 0.7 \\
Companion 7:1   &  13\%             & 0.7 \\
Companion 10:1  &  16\%             & 0.8 \\
\hline
\end{tabular}
\caption{Composition of the galaxy models: amount of gas and dark matter, as a function of the stellar mass.}
\label{params}
\end{table*}

Several parameters describe the galactic encounter:
\begin{itemize}
\item the mass ratio from 10:1 to 1:1, that is the ratio between the {\it stellar disk masses}
\item the direction of the orbit (prograde or retrograde) with respect to the sense of rotation of the most massive galaxy. We only study prograde-prograde and retrograde-retrograde encounters, i.e. the orbit is prograde for the two galaxies, or retrograde for the two galaxies. Prograde-retrograde encounters are not considered in this paper.
\item the impact parameter $r$
\item the relative velocity $V$ of the two galaxies at an infinite distance (the velocity at the beginning of the simulations is inferred from it by neglecting the dynamical friction at large distances)
\item the inclination of each disk with respect to the orbital plane, $i$ for the most massive galaxy and $i'$ for the smaller one
\item the angle $\alpha$ between the two disks
\end{itemize}

We fixed $\alpha=i'=33$ degrees (the mean statistical value in spherical geometry). We used impact parameters $r$ of 18, 35 and 65 kpc, relative velocities at an infinite distance $V$ of 50, 100 and 180 km~s$^{-1}$, and inclination of the orbital plane with respect to the main galaxy disk of 17, 33 and 60 degrees. The values of these parameters in each simulation are given in Table~\ref{runs}. We let the merger remnants evolve for about 10 dynamical times before analyzing their properties.

We stress that as per our definition, the galaxy mass ratio used is the ratio of the stellar masses. The total mass ratios (including gas and dark matter) would be slightly different: since we have assumed that smaller galaxies contain more gas and dark matter, our 10:1 mergers correspond to total mass ratios between 8:1 and 9:1. There could thus be small differences to some papers in the literature that use the total mass ratio. Also, works on minor mergers sometimes neglect the dark halo of the small companion, or implicitly include it in the ''stellar'' mass. This is the case, for instance, in Walker et al. (1996): they study 10:1 mergers, where 10 is the ratio of the stellar masses. Their main galaxy contains dark matter, while the small companion does not. In our study, the small companion contains dark matter (which is more realistic), so that the 10:1 companions will have larger effects. In other words, the 10:1 mergers studied by Walker et al. (1996) would correspond to something like 20:1 with our definition.

\begin{table*}[!ht]
\centering
\begin{tabular}{ccccccccccccc}

\hline
\hline
Run & \multicolumn{4}{c}{Parameters} && \multicolumn{6}{c}{Results} \\
No. & $M$ & Orient & $i$ & $r$ & $V_{\infty}$ &&  Type & $E$ & $t$ & $B/T$ & $v/\sigma_{\|}$ \\ 
\hline

1&10 & P &33&35& 50  && D & 7.1 & 2.5 &0.26& 1.32 \\
2&10 & R &33&35& 50  && D & 6.6 & 2.8 &0.24& 1.53 \\
3&10 & P &33&35& 100 && D & 7.5 & 2.9 &0.23& 1.82 \\
4&10 & R &33&35& 100 && D & 6.9 & 3.1 &0.22& 1.61 \\
5&10 & P &33&18& 100 && D & 7.0 & 2.6 &0.24& 1.77 \\
6&10 & R &33&18& 100 && D & 6.4 & 2.9 &0.26& 1.68 \\
7&10 & P &33&65& 100 && D & 7.8 & 3.0 &0.21& 1.96 \\
8&10 & R &33&65& 100 && D & 7.3 & 3.3 &0.23& 2.04 \\
&&&&&&&&&&&\\
9 &7  & P &33&35& 50  && D & 6.6 & 1.9 &0.36& 1.19 \\
10&7  & P &33&18& 50  && D & 6.3 & 1.7 &0.38& 1.14 \\
11&7  & R &33&35& 50  && D & 6.4 & 2.2 &0.35& 1.27 \\
12&7  & R &66&35& 50  && D & 6.6 & 2.3 &0.34& 1.22 \\
13&7  & P &33&35& 100 && D & 6.6 & 2.1 &0.33& 1.25 \\
14&7  & R &33&35& 100 && D & 6.4 & 2.4 &0.36& 1.38 \\
15&7  & P &33&18& 100 && D & 6.5 & 2.0 &0.32& 1.50 \\
16&7  & R &33&18& 100 && D & 6.2 & 2.2 &0.36& 1.13 \\
17&7  & P &33&65& 100 && D & 6.9 & 2.5 &0.31& 1.74 \\
18&7  & R &33&65& 100 && D & 6.7 & 2.5 &0.33& 1.58 \\
&&&&&&&&&&&\\
19&4.5& P &33&35& 50  && D & 5.6 & 1.1 &0.41& 1.05 \\
20&4.5& P &17&35& 50  && D & 5.3 & 1.1 &0.42& 1.02 \\
21&4.5& P &66&35& 50  && D & 5.8 & 1.0 &0.40& 1.06 \\
22&4.5& P &33&65& 50  && D & 5.4 & 1.4 &0.38& 1.09 \\
23&4.5& P &33&18& 50  && E & 5.7 & 1.2 & -- & 1.02 \\
24&4.5& R &33&35& 50  && D & 5.5 & 1.3 &0.42& 0.95 \\
25&4.5& R &33&65& 50  && D & 5.3 & 1.6 &0.39& 1.17 \\
26&4.5& R &33&18& 50  && E & 5.5 & 1.3 & -- & 0.96 \\
27&4.5& P &33&35& 100 && D & 6.0 & 2.3 &0.39& 0.88 \\
28&4.5& P &33&18& 100 && E & 5.2 & 1.4 & -- & 1.06 \\
29&4.5& P &33&18& 180 && E & 5.3 & 1.5 & -- & 1.05 \\
30&4.5& P &33&65& 100 && D & 6.1 & 1.6 &0.37& 1.11 \\
31&4.5& R &33&65& 100 && D & 5.6 & 1.6 &0.40& 1.26 \\
32&4.5& P &33&35& 180 && D & 6.3 & 1.8 &0.39& 0.89 \\
33&4.5& R &33&35& 100 && D & 5.8 & 2.0 &0.42& 0.78 \\
34&4.5& R &33&18& 180 && D & 5.8 & 1.8 &0.40& 1.17 \\
35&4.5& R &33&18& 100 && E & 5.5 & 1.7 & -- & 0.85 \\
&&&&&&&&&&&\\
36&3  & P &33&35& 50  && D & 4.2 & 0.7 &0.51& 0.90 \\
37&3  & P &66&35& 50  && D & 5.1 & 0.8 &0.62& 0.87 \\
38&3  & R &33&35& 50  && E & 4.0 & 1.1 & -- & 0.85 \\
39&3  & R &17&35& 50  && E & 6.3 & 1.0 & -- & 0.69 \\
40&3  & P &33&35& 100 && E & 5.2 & 1.3 & -- & 0.71 \\
41&3  & R &33&35& 100 && E & 4.9 & 1.6 & -- & 0.64 \\
42&3  & P &33&65& 100 && E & 5.5 & 1.5 & -- & 0.81 \\
43&3  & R &33&65& 100 && E & 5.2 & 1.6 & -- & 0.58 \\
44&3  & P &33&18& 100 && E & 5.1 & 1.2 & -- & 0.92 \\
45&3  & R &33&18& 100 && E & 4.8 & 1.4 & -- & 0.77 \\
46&3  & P &33&18& 50  && E & 4.5 & 0.7 & -- & 0.82 \\
47&3  & P &33&65& 50  && D & 4.7 & 1.1 &0.55& 0.87 \\
&&&&&&&&&&&\\
48&2  & P &33&35& 50  && E & 4.5 &0.6  & -- & 0.67 \\
49&2  & R &33&35& 50  && E & 3.5 &0.8  & -- & 0.42 \\
50&2  & P &33&18& 180 && E & 4.1 & 1.1 & -- & 0.52 \\
51&2  & P &33&35& 100 && E & 4.9 & 0.9 & -- & 0.41 \\
52&2  & R &33&35& 100 && E & 4.2 & 1.2 & -- & 0.29 \\
53&2  & P &33&65& 100 && E & 5.2 & 1.1 & -- & 0.50 \\
54&2  & R &33&65& 100 && E & 4.4 & 1.5 & -- & 0.47 \\
55&2  & P &33&18& 100 && E & 4.4 & 0.9 & -- & 0.71 \\
56&2  & R &33&18& 100 && E & 4.5 & 1.2 & -- & 0.38 \\

\end{tabular}
\end{table*}

\begin{table*}
\centering

\begin{tabular}{ccccccccccccc}
57&1  & P &33&35& 100 && E & 4.4 & 0.7 & -- & 0.28 \\
58&1  & R &33&35& 100 && E & 4.6 & 0.9 & -- & 0.17 \\
59&1  & P &33&65& 100 && E & 4.6 & 0.8 & -- & 0.14 \\
60&1  & R &33&65& 100 && E & 5.0 & 1.1 & -- & 0.09 \\
61&1  & P &33&18& 100 && E & 4.8 & 0.8 & -- & 0.23 \\
62&1  & R &33&18& 100 && E & 4.6 & 0.8 & -- & 0.21 \\
&&&&&&&&&&&\\
C1& \multicolumn{5}{c}{control run} && D & 8.4 & --  &0.20& 3.10 \\
C2& \multicolumn{5}{c}{control run} && D & 8.4 & --  &0.22& 3.08 \\
C3& \multicolumn{5}{c}{control run} && D & 8.6 & --  &0.19& 3.17 \\
\hline
\end{tabular}

\caption{Run parameters and results. See text for the definition of the parameters and properties of the relaxed remnants. Control run C1 is for the same galaxy as the most massive galaxy in the simulations of merger. In control run C2, the initial bulge-to-total mass ratio is 0.19 instead of 0.17. In run C3, its initial value is 0.15.}
\label{runs}

\end{table*}

\subsection{Analysis of the merger remnant properties}\label{defs}

\begin{figure}
\centering
\resizebox{8cm}{!}{\includegraphics{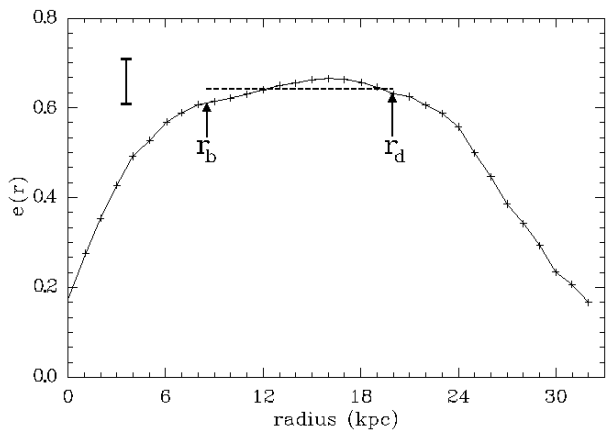}}
\caption{Evolution of $e=10 (1-b/a)$, where $b/a$ is the isophotal axis ratio, versus radius (defined as the distance along the apparent major axis, i.e. the projection of the disk plane), in the relaxed remnant of Run~11, a disk galaxy, observed edge-on. The bulge radius and disk radius (25th isophote) are indicated, as well as the mean value of $e$ between these two radii. The error bar indicated on the figure corresponds to the variations of $e(r)$ between different edge-on projections. This physical uncertainty dominates the statistical error on the measure of $e(r)$ for a given projection. We give the average uncertainty ; there is no strong variation of it with radius.}
\label{er}
\end{figure}

\begin{figure}
\centering
\resizebox{8cm}{!}{\includegraphics{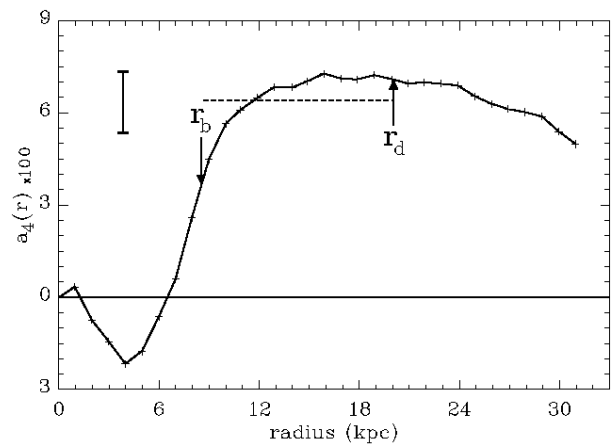}}
\caption{Evolution of the diskiness parameter $a_4$ versus radius, in the relaxed remnant of Run~11. We have derived the mean value of $a_4$ over four different edge-on projections (see text for details). The bulge radius and disk radius (25th isophote) are indicated, as well as the mean value of $a_4$ between these two radii. Note that the bulge is boxy ($a_4<0$) in this disk galaxy. The error bar shown in the figure has the same meaning as the one shown in Fig.~\ref{er}.}
\label{a4r}
\end{figure}

Several properties have been computed for the merger remnants, 4~Gyr after the beginning of the simulations, i.e. when they are fully relaxed:

\begin{itemize}
\item its morphological type, either {\it elliptical} or {\it disk} galaxy. When an exponential radial luminosity profile is found (except in the central part where a bulge is present), the system is classified as a disk remnant -- we considered an exponential fit $\exp^{-r/r_e}$ as robust, and sufficient to classify the system as a disk galaxy, when the luminosity profile is fitted over a radial range larger than $1.5 r_e$; this criterion is discussed and justified in Sect.~3.1. When no robust exponential fit is found on the luminosity profile, the system is classified as an elliptical galaxy. We check a posteriori that all these elliptical class remnants follow the de Vaucouleur (1997) $r^{1/4}$ law, but we do not assume a priori that the $r^{1/4}$ profile is characteristic of all elliptical galaxies. Also, when a exponential disk is found, the bulge is regarded as the central excess of luminosity compared to the exponential component, but no assumption is made as to what the luminosity profile of a bulge should be. 
\item for remnants classified as disk galaxies, we measured the bulges properties: its mass, and its extent (or bulge radius), which is the radius at which the deviation from the exponential profile becomes negligible. A similar definition of the bulge radius has been adopted by L\"utticke et al. (2004). In this paper, we often use the bulge-to-total mass ratio, which is the ratio of the bulge mass to the total visible (disk+bulge) mass. It is smaller than the bulge-to-disk mass ratio. To compute the bulge mass, we subtract the disk contribution in the inner regions, by extrapolating the exponential profile in the inner regions.
\item for all the remnants, we measured the ellipticity parameter $e(r)$ and the diskiness parameter $a_4(r)$, as a function of the radius $r$. We used the same definition for these parameters as Naab \& Burkert (2003): $e(r)$ is $10\times(1-b/a)$ where $b/a$ is the isophotal axis ratio in an ellipse-fitting model. $a_4(r)$ is the coefficient of $\cos(4 \phi)$ in the Fourier expansion of the isophotal shape, $a_4$ is positive for disky isophotes, and negative for disky isophotes. We measured $a_4(r)$ and $e(r)$ for ''edge-on'' orientations: the ''face-on'' projection has been assumed to be the one for which the 25th isophote in an ellipse-fitting model was circular. Several ''edge-on'' projections were then possible, so we analyzed ten projections, with azimuthal rotations of 18 degrees between each one. We then computed for each radius $r$ the mean values of $e(r)$ and $a_4(r)$ over these four projections. Then, we wanted to keep a single value for these parameters, to enable a simple comparison between the simulated merger remnants. For remnants classified as disk galaxies, the curves of $e(r)$ and $a_4(r)$ are generally flat over a large radial range between the bulge radius and the disk optical radius (25th isophote), or even further, as shown for instance by Figs.~\ref{er} and \ref{a4r} 
for Run~11. 
We thus defined $a_4$ and $E$ as the mean values of $a_4(r)$ and $e(r)$ over this radial range. For elliptical remnants, we derived them as the mean values over the range $\left[ 0.55 R_{25} ; R_{25} \right]$. Since the bulge extent in the more bulge-dominated disk galaxies in our sample is about $0.55 R_{25}$ (See Sect.~4), this ensures the continuity of the analysis between the disk and elliptical remnants, so that it makes sense to compare the values of $E$ and $a_4$ between these two types of remnants. We checked that the above choice does not result in any significant difference to the values given by Naab \& Burket (2003) in their analysis of major merger remnants: they used a different radial range to compute their mean values, but for elliptical-like remnants, $a_4(r)$ and $e(r)$ are rather constant over a range larger than $\left[ 0.55 R_{25} ; R_{25} \right]$.
\item to describe the global kinematical properties of the merger remnants, we measured $v/\sigma$ along the main plane corresponding to the ''face-on'' projection defined above (disk plane for a disky remnant) and perpendicular to it. We then computed the mean values over the radial range used for $a_4$ and $E$. As shown by Fig.~\ref{kin}, $v/\sigma$ is rather constant over this radial range.
\item we also define the merging time as the time at which the distance between the mass center of the two systems becomes smaller than 5 kpc. The choice of this distance is not crucial, using 2 or 10 kpc instead of 5 does not result in major changes.
\end{itemize}

We checked in several cases that the comparison of the mean values $a_4$, $E$ and $v/\sigma$ between two systems give the same result as the comparison of the whole profile $a_4(r)$, $e(r)$, and $v/\sigma (r)$. The ellipse-fitting of the isophotes and the computation of $a_4(r)$ and $e(r)$ were made using the \texttt{stsdas} package in IRAF.

We have also run a control simulation of the same main galaxy, evolving as an isolated system over the same period. This enables us to see which part of the evolution of $a_4$, $E$, $v/\sigma$ or the bulge mass is caused by the merger, and which part is related to secular evolution.

\section{Disky and elliptical merger remnants}

\subsection{Morphology}

\subsubsection{Luminosity profile of the merger remnants}

\begin{figure*}
\centering
\resizebox{11.8cm}{!}{\includegraphics{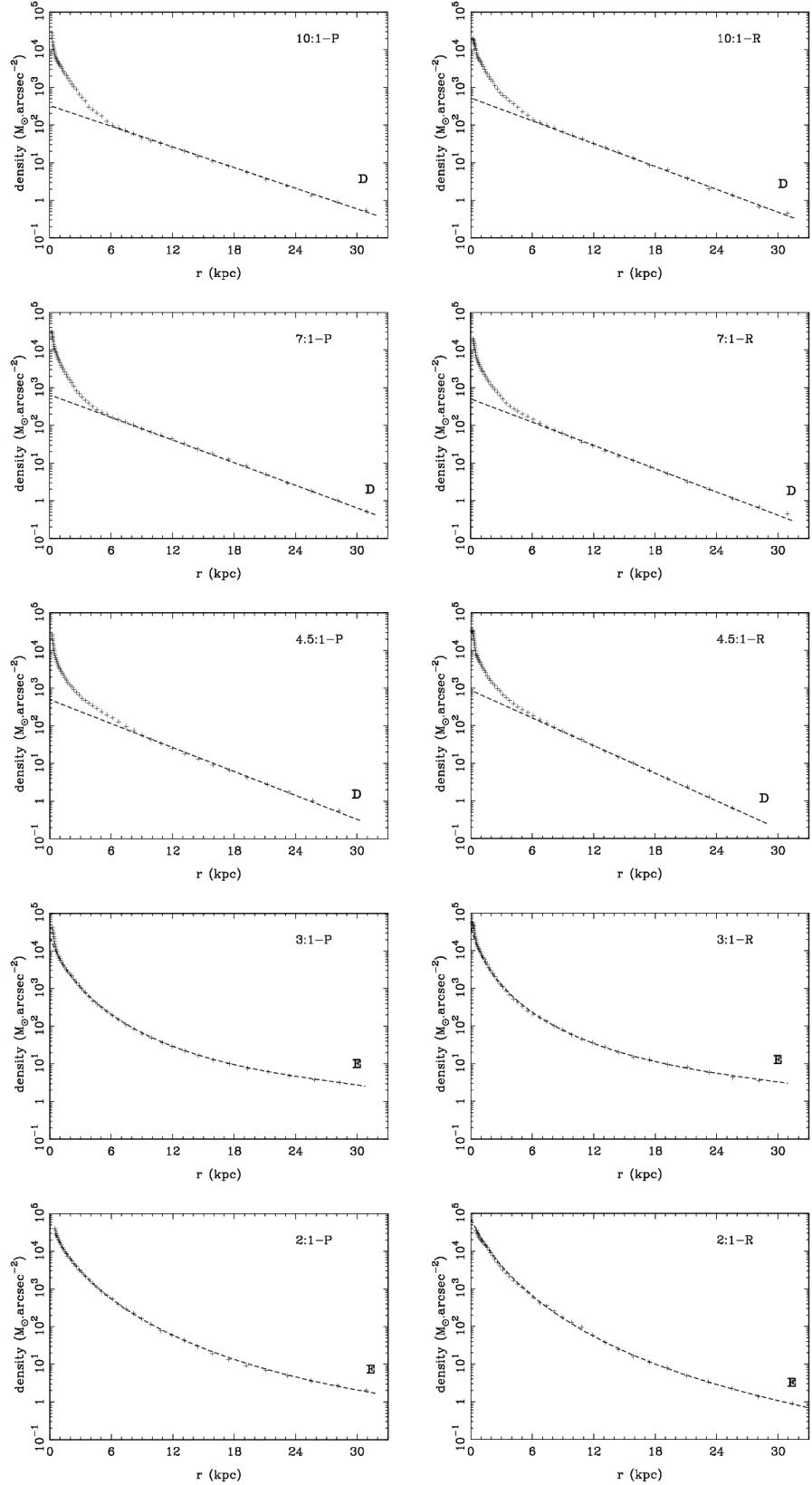}}
\caption{Radial luminosity profiles for a set of galaxy merger remnants, with mass ratios ranging from 10:1 to 2:1, prograde orbits (left column) and retrograde orbits (right column). For all these runs, $V=100$~km~s$^{-1}$, $r=35$ kpc, and $i=33$ degrees. Exponential or de Vaucouleurs fits have been plotted, depending on whether the system is classified as a disk or elliptical galaxy, according to the classification criterion detailed in the text. The labels 'D' or 'E' on each profile correspond to this classification.}
\label{profiles}
\end{figure*}

Our first purpose is to classify the merger remnants between elliptical galaxies and disk galaxies. We analyze the relaxed systems as if they were observed ''face-on'': we choose the projection that makes the outer isophote circular. The azimuthally averaged luminosity profiles of several merger remnants are shown in Fig.~\ref{profiles}. The 10:1, 7:1 and 4.5:1 remnants seen in this figure show an exponential disk and a central bulge. The bulge is much more massive than before the merger or in the control run (see Table~\ref{runs}), but the mass distribution is still dominated by the exponential disk component. The mass distribution of these merger remnants is therefore similar to an early-type spiral galaxy. We consider that the this disk component is extended enough to be detected when an exponential fit can be found to the face-on luminosity profile over a radial range $1.5$ times larger than the exponential scale-length: this choice is discussed below. Whether the disk can be detected under different orientations will be discussed later. Note that this criterion for classifying a merger remnant as a ''disk'' galaxy does not require any fit to the bulge profile, so no assumption for the bulge profile has to be made. 

\begin{figure*}
\centering
\resizebox{7.9cm}{!}{\includegraphics{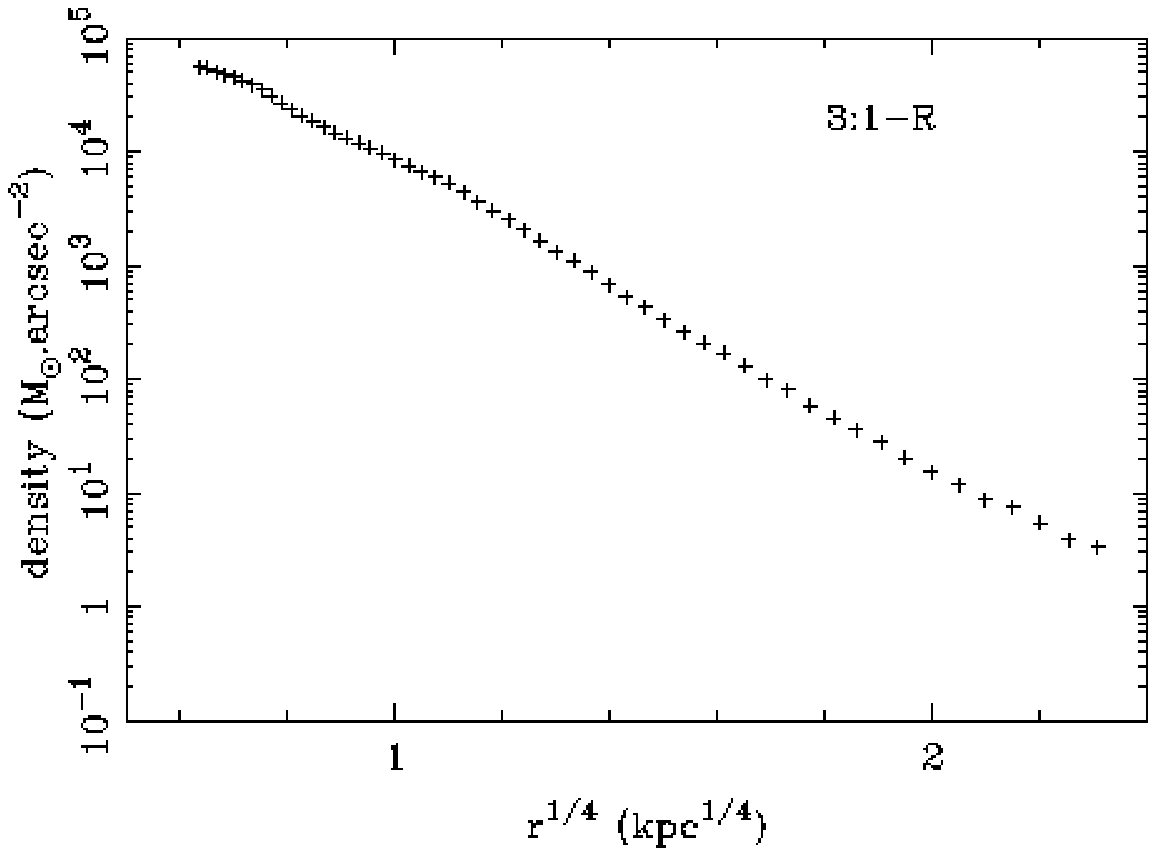}}
\resizebox{7.9cm}{!}{\includegraphics{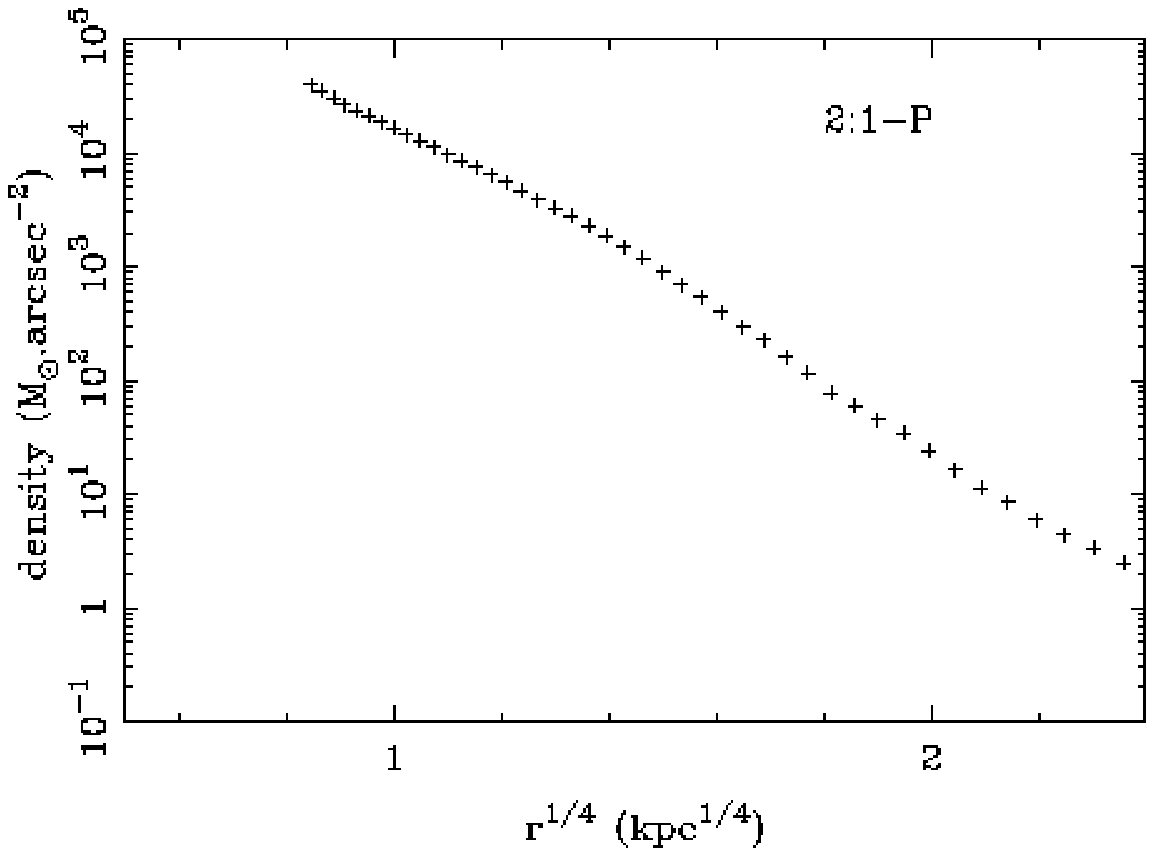}}
\caption{Radial luminosity profiles of two runs with elliptical-like remnants (3:1 retrograde orbit (Run~41), and 2:1 prograde orbit (Run~51), with $V=100$~km~s$^{-1}$, $r=35$~kpc, $i=33$ degrees). The linear aspect in this $r^{1/4}$--magnitude frame shows that the radial luminosity profile can be well-fitted by a de Vaucouleurs (1977) profile. See Fig.~3 for the corresponding radial profiles and $r^{1/4}$ fit.}
\label{profilesE}
\end{figure*}

At the opposite end of mass ratios, the luminosity profiles of the 3:1 to 1:1 merger remnants displayed in Fig.~\ref{profiles} do not show any robust exponential component: a poor exponential fit is only possible in the outer regions. Then, we cannot classify them as ''disk'' galaxies, and instead call them ''elliptical'' remnants. We verify a posteriori that their luminosity profiles can be well fitted by a $r^{1/4}$ profile, as shown in Fig.~\ref{profilesE} for two cases, even if Sersic profiles with index $n\neq 4$ may provide better fits -- however, the Sersic index of elliptical systems is beyond the scope of this paper; we only want to separate the remnants into disk galaxies and elliptical systems.

\begin{figure*}
\centering
\resizebox{15.8cm}{!}{\includegraphics{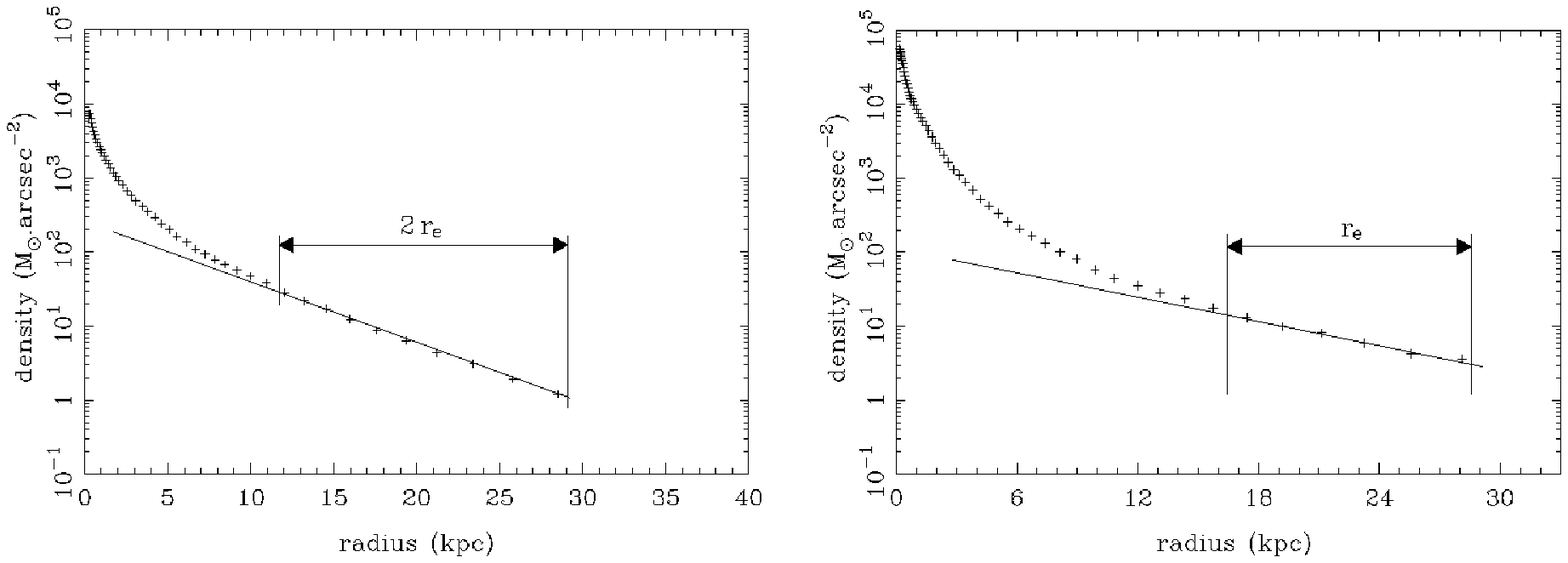}}
\caption{Illustration of an exponential disk of profile $\exp ^{-r/r_\mathrm{e}}$ fitted over a range $\Delta r \simeq 2 r_\mathrm{e}$ (left): the disk detection looks robust. Right: a case where the best exponential fit corresponds to $\Delta r \simeq r_\mathrm{e}$. The exponential fit does not seem robust, and on the contrary this system is well-fitted by an $r^{1/4}$ profile (see the same system on the left panel of Fig.2). On the basis of such examination of luminosity profiles, we fixed $\Delta r \simeq 1.5 r_\mathrm{e}$ as the limit for an exponential disk detection to be considered robust.}
\label{Dr}
\end{figure*}

We made a similar analysis for each relaxed merger remnant. The results are given in Table~\ref{runs}. At this stage, we classified a remnant as a ''disk'' galaxy if an exponential profile $\exp ^{-r/r_\mathrm{e}}$ can be fitted over a radial range $\Delta r$ as large as at least $1.5 r_\mathrm{e}$ inside the 25th isophote. The reasons for this choice are:
\begin{itemize}
\item an examination of many profiles has shown that an exponential fit is very robust when established over a radial range $\Delta r=2 r_\mathrm{e}$, still looks rather robust when $\Delta r = 1.5 r_\mathrm{e}$, but is very poor and not reliable when $\Delta r = r_\mathrm{e}$ (see Fig.~\ref{Dr}).
\item in their observational study, Chitre \& Jog (2002) have been able to detect disks with $\Delta r = 1.5-2 r_\mathrm{e}$ or larger, but not smaller.
\end{itemize}
As shown in the next Section, the morphological classification resulting from this criterion is compatible with a classification that would be based on the vertical mass distribution, suggesting that we chose the correct criterion to classify galaxies either as ''disks'' or as ''ellipticals''.

All the systems that are not classified as ''disk'' galaxies, because of a very poor exponential fit, have been classified as ''elliptical'' remnants --  we have checked for each of them that their radial distribution can be well fitted by a $r^{1/4}$ profile, as is the case for the two systems shown in Fig.~\ref{profilesE}. This concerns 1:1 and 2:1 remnants, most 3:1 cases and a few 4.5:1 cases.

\subsubsection{Visibility of disks}

The main concern with the morphological classification of the merger remnants established above is that observationally, disks may be missed. We said before that the $\Delta r \geq 1.5 r_\mathrm{e}$ criterion selects robust fits for which the disk component is rather obvious and may not be missed observationally, which is true for the face-on systems that we have studied so far. But when the system is not observed face-on, the disk profile is not purely exponential any longer (even if not largely different from exponential), and the range over which it can be fitted is smaller (the bulge may hide a part of the disk).

\begin{figure}
\centering
\resizebox{8cm}{!}{\includegraphics{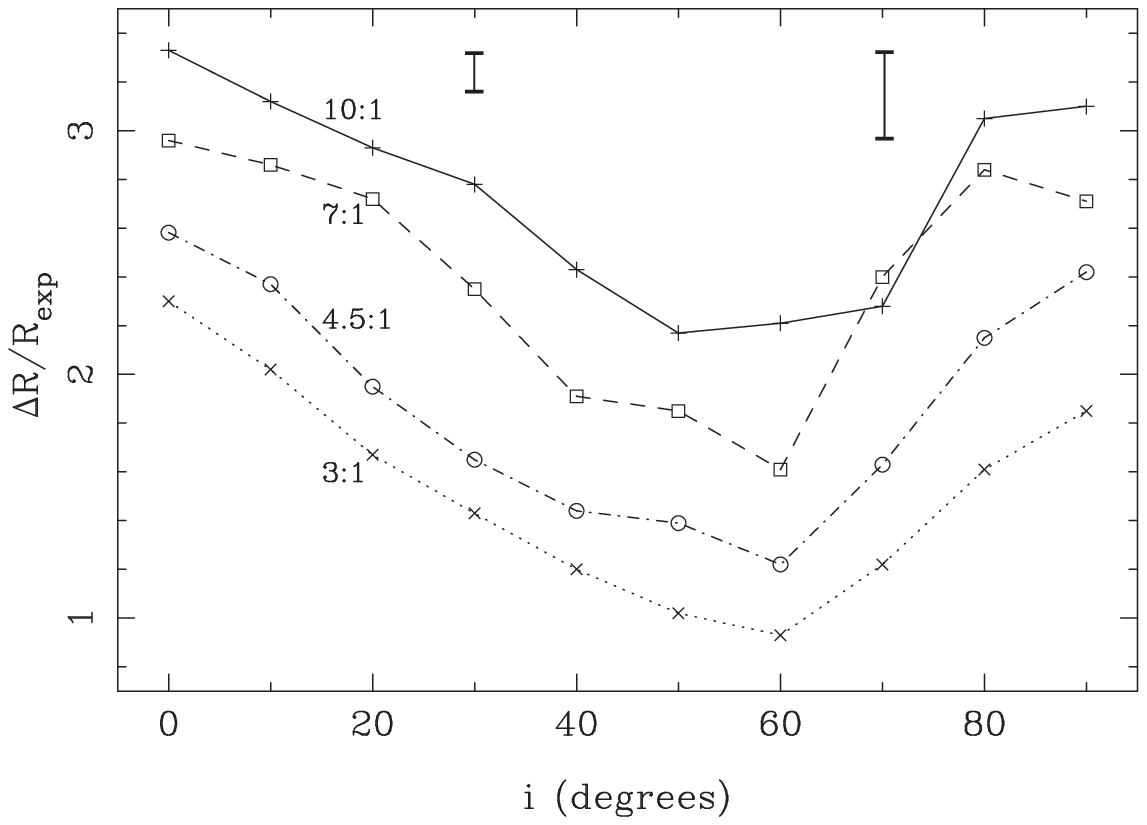}}
\caption{Radial range $\Delta R$ over which the theoretical profile of a disk can be fitted to the actual profile, compared to the exponential scalelength of this disk, $r_\mathrm{e}$, as a function of the inclination of the line-of-sight. We argue in the text that the threshold for the disk detection should be $\Delta R \geq 1.5 r_\mathrm{e}$. Disks in the 4.5:1--10:1 remnants are then detected most of the time. Measurements have been made on runs with $V=50$~km~s$^{-1}$, $r=35$~kpc, $i=33$ degrees. We have computed average values for prograde and retrograde orbits (except for the 3:1 case, where only the prograde remnant shows a disk), and for 10 different projections for each value of the inclination. The bulge contamination, resulting from the visual projection of the bulge light on the disk, is maximum for the inclination of 50--70 degrees, for which the disk is the least visible. The error bars represent the average uncertainty due to the fact that several projections are possible for a given value of $i$ (except for $i=0$). We show it for $i=30$ and 70 degrees.}
\label{visi}
\end{figure}

\paragraph{Luminosity profile for random inclination}
We have analyzed the luminosity profiles of merger remnants with several inclinations, and show in Fig.~\ref{visi} the range over which a disk component can be fitted $\Delta r$, compared to the exponential scalelength $r_\mathrm{e}$, as a function of the inclination $i$ of the system. We said before that a likely limit for the detection of disks is $\Delta r \geq 1.5 r_\mathrm{e}$. According to this, the disk components of the 7:1 and 10:1 merger remnants should be detected whatever the inclination. For the 4.5:1 merger remnants, the disk can be missed if $40<i<65$. Since the probability of an inclination $i$ is proportional to $\sin(i)$, this means that, in a sample of 4.5:1 merger remnants with random orientations, only 28\% of the disks in the 4.5:1 merger remnants are likely to be missed.

A few 3:1 remnants (only with small $V$ and $r$, see Table~2) may show a robust exponential disk component when observed face-on, but for random orientation, the fit is generally poor, and the system is likely not to be classified as a disk galaxy: according to the detection criterion above, the probability that the disk is missed in these 3:1 merger remnants is 62\%, but this is not a serious constraint on the detection of disks because for this mass ratio, a disk results for only a few cases.

	\paragraph{Vertical mass distribution}

\begin{figure*}
\centering
\resizebox{12cm}{!}{\includegraphics{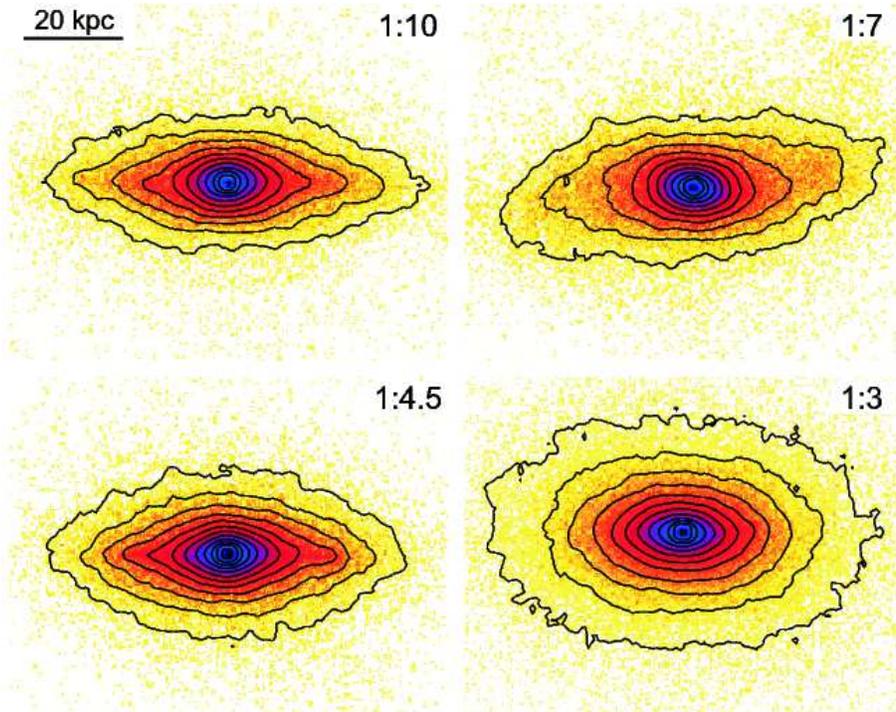}}
\caption{Edge-on density maps of four merger remnants with various mass ratios. Parameters: $V=50$~km~s$^{-1}$, $r=35$~kpc, $i=33$ degrees, prograde orbit for the 10:1 (Run~1) and 4.5:1 (Run~19) cases, and retrograde orbits for the 7:1 (Run~11) and 3:1 (Run~38) cases. Note the high diskiness of the merger remnant (at radii $\sim$ 10kpc) in the 4.5:1--10:1 range of mass ratio, showing the presence of the disk, as inferred from their radial mass distribution. Also note, in the 7:1 merger remnants, the boxiness of the bulge, at radii $\sim$5 kpc.}
\label{maps}
\end{figure*}

\begin{figure}
\centering
\resizebox{8cm}{!}{\includegraphics{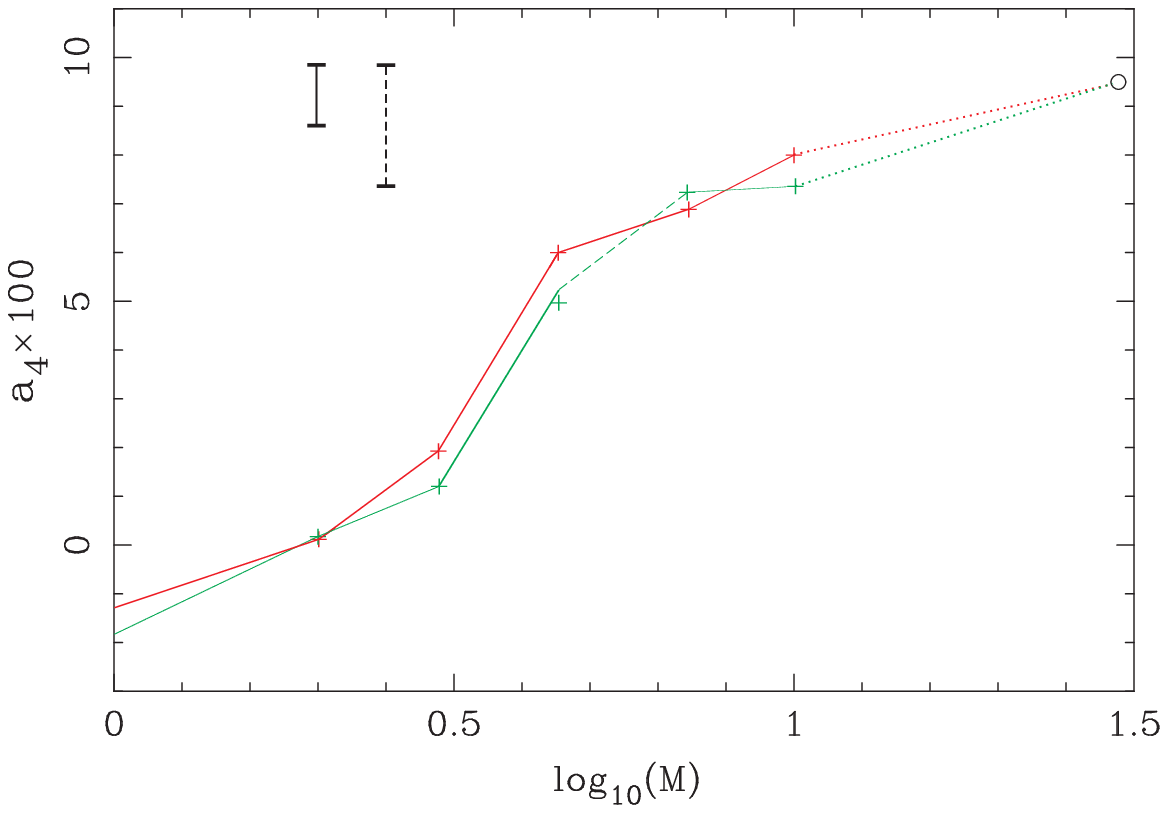}}
\caption{Diskiness parameter $a_4$ as a function of the mass ratio M. Red solid curve: $V=50$~km~s$^{-1}$, $r=35$~kpc, $i=33$ degrees, retrograde orbit -- Green dashed curve: $V=100$~km~s$^{-1}$, $r=35$~kpc, $i=33$ degrees, prograde orbit. We show mean values over 4 different edge-on projections. Black circle: value for the control run (isolated galaxy with only secular evolution). Note that the disk remnants in the 4.5:1--10:1 range of mass ratios are significantly more disky than the elliptical galaxies with disky isophotes resulting from the 3:1 mergers (also called ''disky ellipticals'', which does not mean that they are disk galaxies). The solid error bar is the uncertainty associated with the different possible edge-on projections, for $a_4$ varies when an azimuthal rotation is applied to the system. This uncertainty is larger than the statistical error on the measure of $a_4$ for a given projection. The dashed error bar corresponds to variations when orbital parameters are varied (this is a real physical variation of $a_4$, not an uncertainty).}
\label{a4}
\end{figure}

Until now, we have not accounted for the presence of dust. Interstellar dust could prevent the detection of disks when they are seen edge-on, for dust absorption in an edge-on system is able to significantly modify their radial luminosity profile when observed in the optical light. Yet, in the case of edge-on systems, there is other evidence for the presence of a disk component:
\begin{itemize}
\item the 10:1 and 7:1 remnants are more flattened than most elliptical galaxies (see Table~\ref{runs}).
\item the isophotes of the disk-like 4.5:1, 7:1 and 10:1 remnants are strongly disky with values of the $a_4$ parameter, as defined in Sect.~2, of 0.05--0.09 (see Fig.~\ref{a4}), much larger than in disky elliptical galaxies\footnote{Disky elliptical galaxies are elliptical galaxies with disky isophotes. This does not mean that they contain a disk: they are much less disky than true disk galaxies, and do not have the same radial distribution. } (see e.g., Naab \& Burkert (2003) for typical values in elliptical galaxies). In these remnants, the isophotes are so disky that even an eye examination can attest to the presence of a massive disk component (see for instance Fig.~\ref{maps}).
\end{itemize}

This will enable observers to detect the disk even in case of strong dust absorption that may disturb or even hide the characteristic profile of an exponential disk. Moreover, the dust itself is an indicator of the presence of a disk in an edge-on system, as already noticed by Rix \& White (1990): as explained in Sect.~\ref{gas}, the 4.5:1, 7:1 and 10:1 remnants can still contain a few percent of gas, so the prominent dust lane characteristic of an edge-on disk will be visible.

Furthermore, the vertical mass distribution in the systems that we have classified as ''disk'' galaxies on the basis of their radial profile, characterized by the values of $a_4$ and $E$ mentioned above and in Table~\ref{runs}, is typical of spiral galaxies. This confirms that we were right in classifying these merger remnants as disk galaxies. It also suggests that the criterion $\Delta r \geq 1.5 r_\mathrm{e}$ for the robustness of an exponential disk is correct, since we did not classify as disks systems that do not have a vertical distribution typical of a disk galaxy. As shown by Fig.~\ref{a4}, there is a clear transition between the (disky) elliptical remnants for 3:1 mergers, and the disk remnants for 4.5:1 mergers, when one examines their vertical mass distribution: disk remnants formed in 4.5:1 mergers are much more disky than the most disky elliptical galaxies.

\paragraph{Conclusion}

We have classified as ''disk'' galaxies systems in which a robust exponential disk can be seen. Other merger remnants have been called ''elliptical'' galaxies, which is justified since we have checked that an $r^{1/4}$ profile provides a good fit to their luminosity profiles. The vertical distribution of matter has been shown to be consistent with this classification.

The 10:1 and 7:1 merger remnants, and most of the 4.5:1 merger remnants, have then been classified as disk galaxies. The exponential disk contains most of the visible mass, even if a massive central bulge is also present. These systems have a significant flattening, and a very disky isophotal shape. Even if the disks of some 4.5:1 remnants could be missed observationally, these merger remnants have morphological properties of early-type disk galaxies. Only a few 4.5:1 cases have resulted in systems that do not have a robust disk, but rather resemble elliptical galaxies: they correspond to the smallest impact parameters, that are also the least likely to occur.

Two 3:1 mergers in our sample have resulted in systems where a disk component is found, but this disk is less massive than the central bulge, and would be difficult to observe when not seen face-on. The range of mass ratios 1:1--3:1 mainly result in galaxies that have no massive exponential disk, but that are well fitted by an $r^{1/4}$ radial profile. The detailed properties of such major merger remnants have already been studied in several works (see references in the Introduction).
Our results regarding their flattening and the diskiness of their isophotes (Table ~\ref{runs} and Fig.~\ref{a4}) are in agreement with these other findings.

The morphological type (disk or elliptical) of the merger remnants is thus mainly dependent on the mass ratio. The influence of other parameters is much less important. We thus conclude that the morphological transition between major mergers, giving birth to elliptical galaxies, and mergers resulting in disturbed, hybrid disk galaxies, occurs in a well defined range of mass ratios, between 3:1 and 4.5:1.

\subsection{Kinematics}

\begin{figure}
\centering
\resizebox{8cm}{!}{\includegraphics{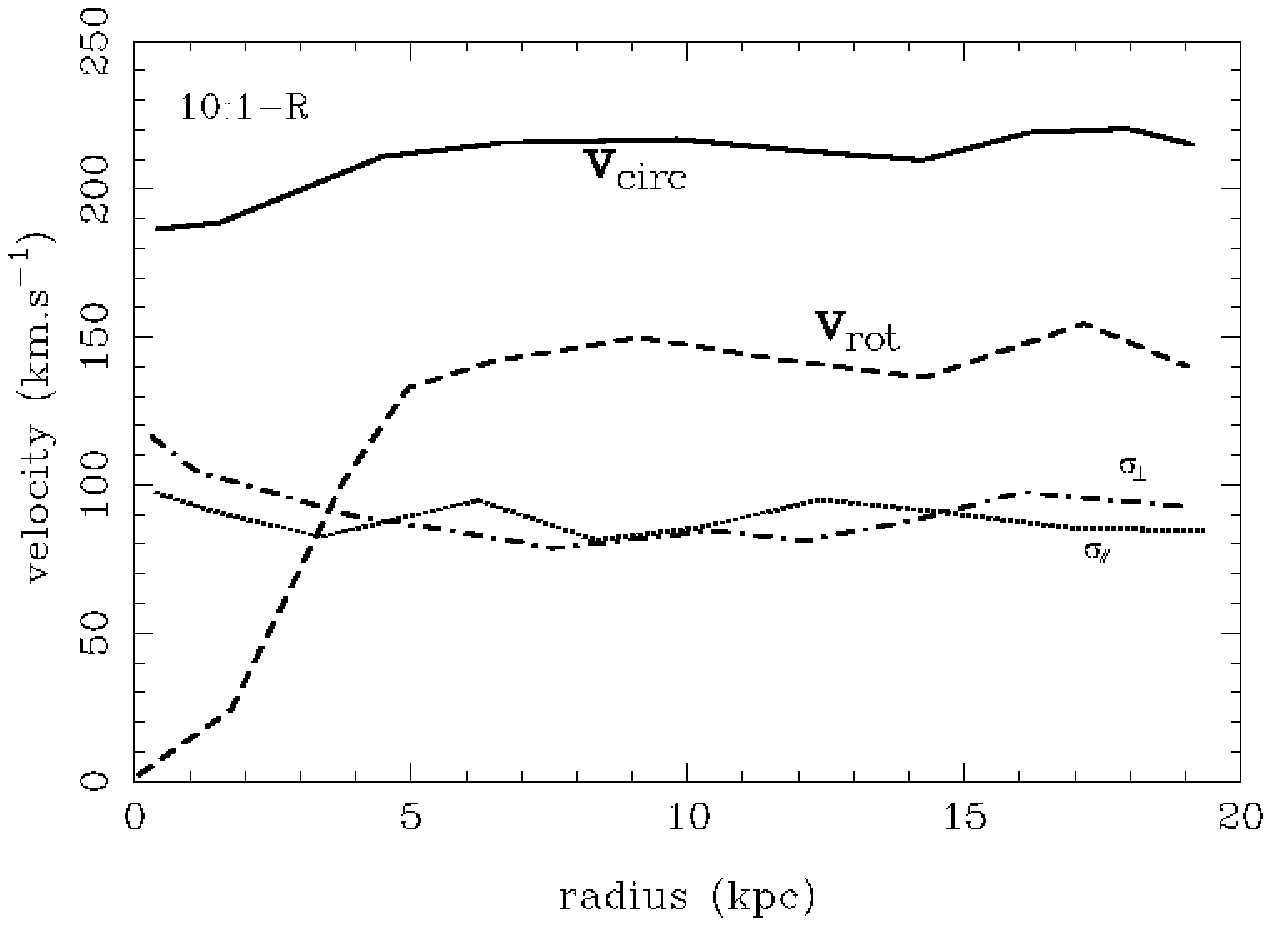}}
\resizebox{8cm}{!}{\includegraphics{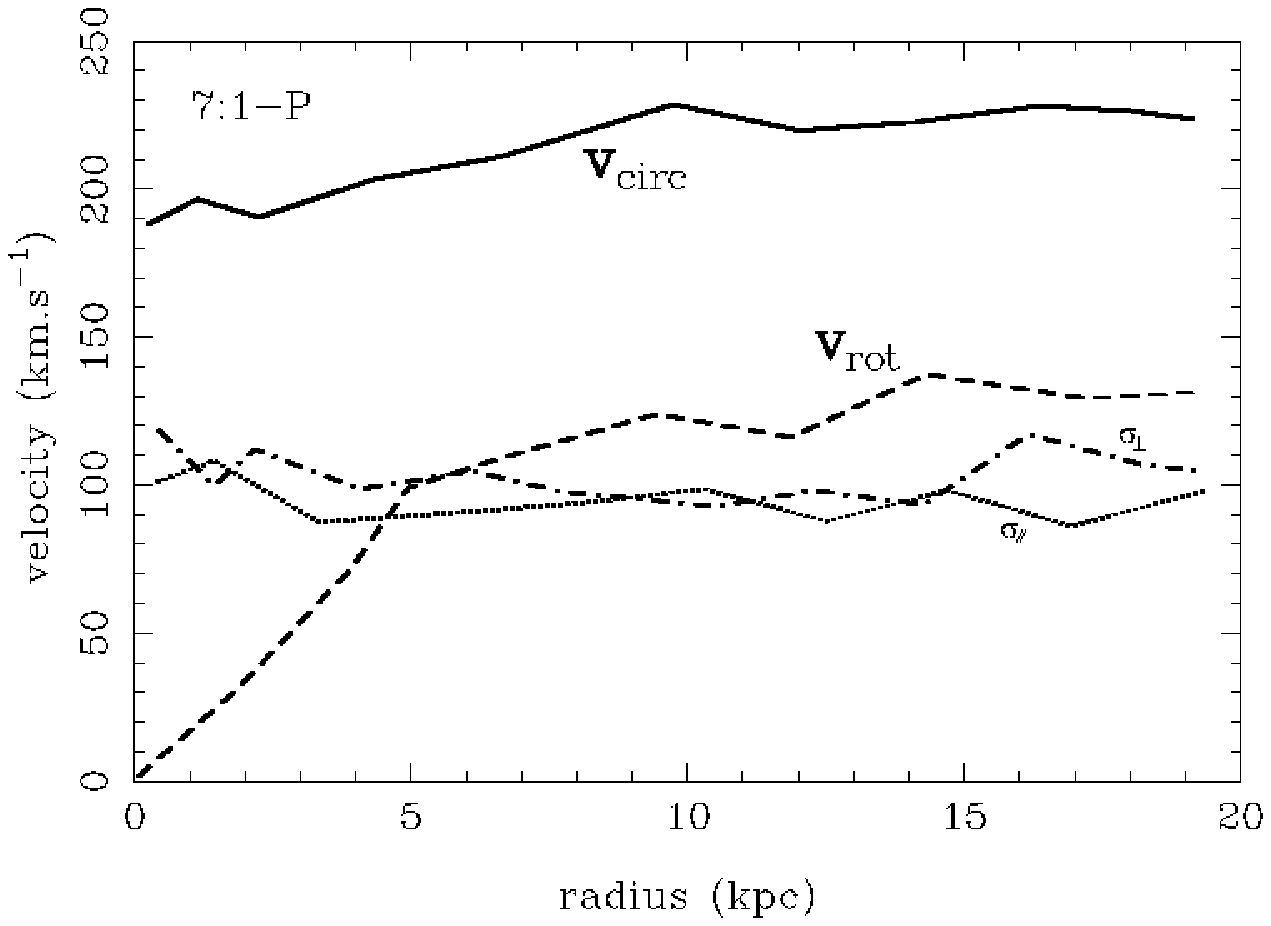}}
\resizebox{8cm}{!}{\includegraphics{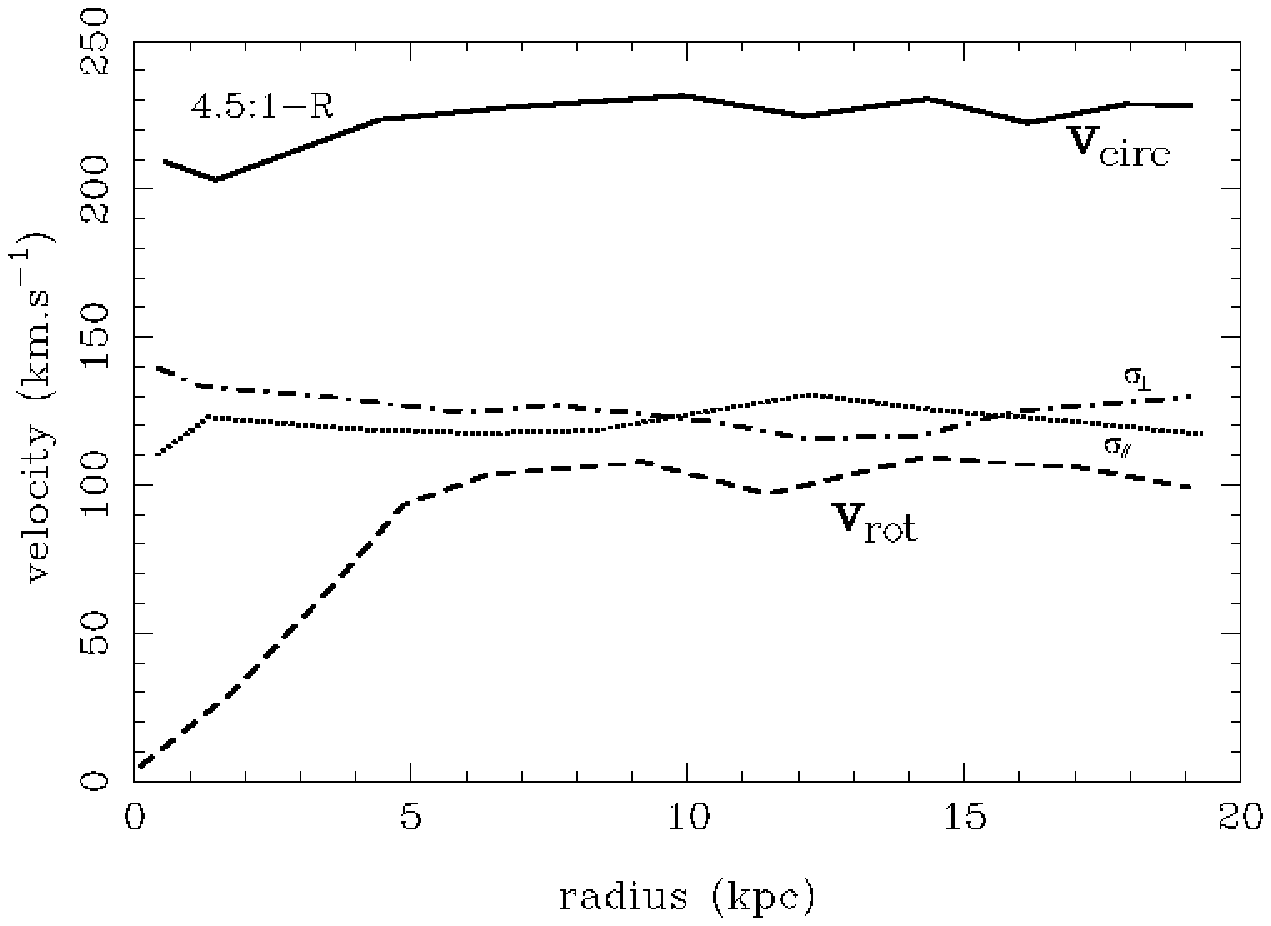}}
\resizebox{8cm}{!}{\includegraphics{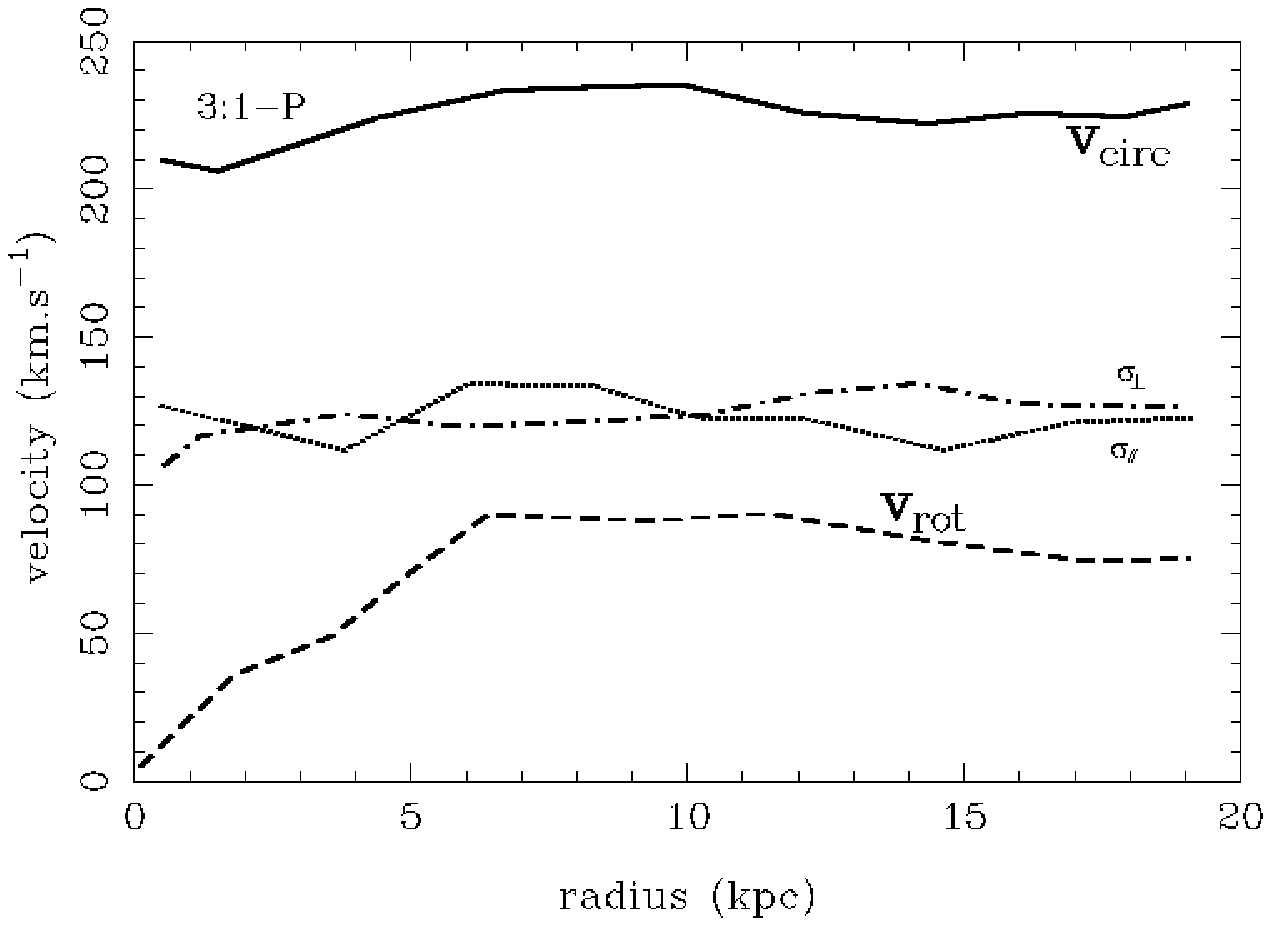}}
\caption{Kinematical profiles of several relaxed remnants with various mass ratios (Runs~4, 13, 27, and 40). Typical uncertainties on the velocities (rotation or dispersion) are 5 to 10\% at $r=5kpc$ and up to 15\% at $r= 20kpc$.}
\label{kin}
\end{figure}

\begin{figure}
\centering
\resizebox{8cm}{!}{\includegraphics{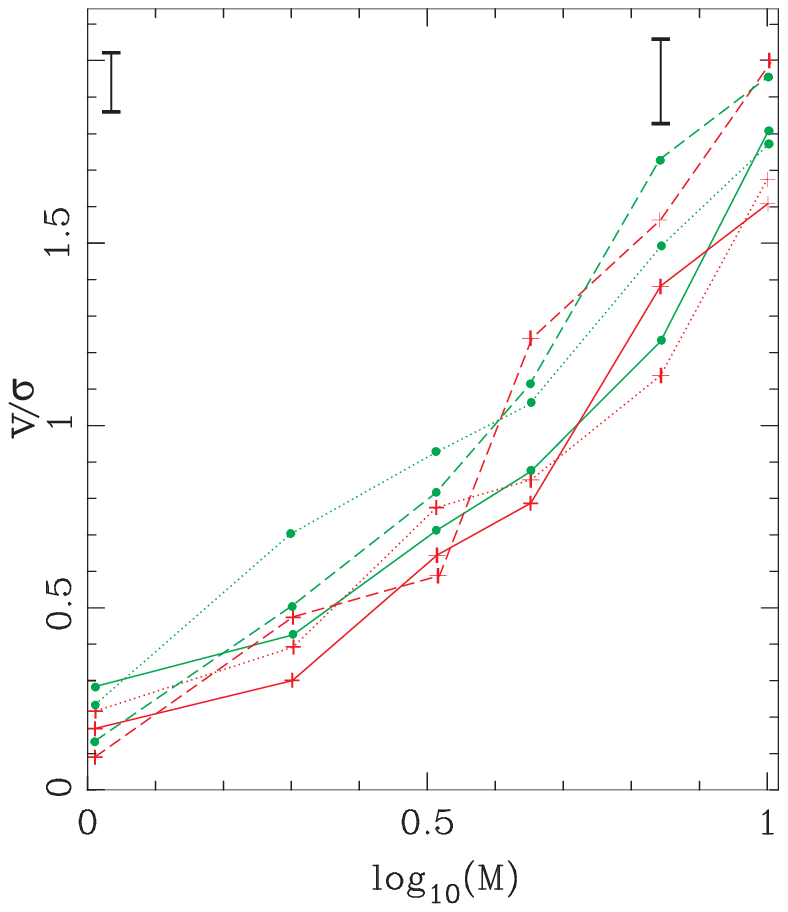}}
\caption{Evolution of $v/\sigma$ along the disk plane vs the mass ratio M from 1:1 t 10:1 (See Sect.~2.3 for the exact definition of $v/\sigma$). Symbols: green lines/circles : prograde orbits , red lines/crosses : retrograde orbits -- dotted lines for $r=18$~kpc, dashed lines for $r=65$~kpc, and solid lines for $r=35$~kpc. We here analyzed mergers with parameters $V=100$~km~s$^{-1}$ and $i=33$ degrees. The error bar is the uncertainty associated to the different possible edge-on projections. We give the average uncertainty for the 6 curves, for mass ratios of 1:1 and 7:1.}
\label{vsig}
\end{figure}

We have computed the rotation velocity $v$ and the velocity dispersion $\sigma$ for the relaxed merger remnants. The mean values of $v/\sigma$, measured as indicated in Sect.~2, are given in Table~\ref{runs}, and the  rotation curves and the dispersion profiles for four cases are given in Fig.~\ref{kin}. We also show in Fig.~\ref{vsig} the variations of $v/\sigma$ with the mass ratio and other parameters.

The values of $v/\sigma$ found for the elliptical remnants in the 1:1--3:1 range of mass ratios are in agreement with other works (e.g., Naab \& Burkert 2003). As shown in Fig.~\ref{vsig}, the 1:1 remnants are slow rotators, that at the same time tend to have boxy isophotes (Fig.~\ref{a4}). At the opposite end of this range, the 3:1 mergers produce disky ellipticals with $v/\sigma = 0.5-0.9$.

The morphological transition between elliptical and disk remnants in the 3:1--4.5:1 range is not associated with a very large change in the mean values of $v/\sigma$ (see Fig.~\ref{vsig}), whereas the morphology shows a sharp change over the same mass range, as indicated by the diskiness of the remnants. The 4.5:1 merger remnants are still kinematically hot systems, with $v/\sigma = 0.7-1.3$. For the 7:1 remnants, we find $v/\sigma = 1.0-1.7$, and $v/\sigma = 1.3-2.15$ for the 10:1 cases. The values of $v/\sigma$ are even smaller if we compute the mean value over the whole system, and not over the disk component alone, as is the case for the values given above. These large velocity dispersions are not an effect of secular evolution, or a numerical artifact, since the control simulation shows $v/\sigma \simeq 3$. Thus, in many of these remnants, the velocity dispersion is as large as or even larger than the rotation velocity. These systems are likely to correspond to the ''hybrid'' merger remnants with spiral-like morphologies but elliptical-like kinematics, observed by Jog \& Chitre (2002), that we have studied in Bournaud et al. (2004).

Thus, merger remnants with mass ratios between 4.5:1 and 10:1 have much larger velocity dispersions than spiral galaxies, even if their morphology is typical of early-type disk galaxies. For mass ratios of 10:1, we find the first systems that are really dominated by rotation, with $v/\sigma \geq 2$. On the other hand, the velocity dispersions in 4.5:1 and 7:1 remnants remain smaller than in typical elliptical galaxies, with $v/\sigma$ close to 1 or even slightly smaller, but not much smaller than 1 as is the case for massive elliptical galaxies -- only very low-mass elliptical galaxies can have $v/\sigma \geq 1$, up to 2 (Cretton et al. 2001). That these hybrid remnants, formed in the 4.5--10:1 mergers, could be S0 galaxies will be discussed later.

\subsection{Summary : Classification of the merger remnants}

The morphological and kinematical criteria described above led us to define three classes of galaxy mergers:

\begin{itemize}
\item the {\it major mergers}, resulting in elliptical galaxies
\item the {\it intermediate mergers}, resulting in disk galaxies with very large velocity dispersions, that are not similar either to elliptical galaxies, because of their radial profile, vertical mass distribution, and gas content (see Sect.~4.3), or to spiral galaxies, because of their kinematics. At this stage we call them ''hybrid'' merger remnants, as in Bournaud et al. (2004). They could be S0 galaxies, as discussed in Sect.~5.
\item the {\it minor mergers}, resulting in disturbed spiral galaxies.
\end{itemize}

The morphological transition between the major and intermediate mergers has been shown to occur in the well-defined range of mass ratios 3--4.5:1. The kinematical transition between intermediate mergers and minor mergers is not well-defined, for the mass ratio is not the only parameter that controls the kinematics of the remnant: for instance, as shown in Fig.~\ref{vsig}, some 10:1 remnants have a larger velocity dispersion than some 7:1 remnants. Yet, since the first systems with $v/\sigma \geq 2$, that can be regarded as ''rotating disks'', in other words as ''disturbed spiral galaxies'', are found for the 10:1 mass ratio, it seems that 10:1 is  representative for the transition between intermediate and minor mergers.


\section{Properties of the disk-like remnants}

We now explore in more detail the properties of the disk galaxies formed in the intermediate 4.5:1--10:1 mergers. Some of them, such as the isophotal shape and disk flattening, have already been described  before.

\subsection{Influence of orbital parameters on the properties of merger remnants}
\paragraph{Morphology}
Our coverage of the parameter space (see Table~\ref{runs}) shows that the disky merger remnants tend both to be thicker and to have a more massive bulge component when:
\begin{itemize}
\item the orbit is retrograde instead of prograde
\item the impact parameter is smaller
\item the encounter velocity is smaller
\item the inclination of the orbital plane with respect to the main galaxy disk is smaller
\end{itemize}

Each of these 4 conditions leads to a large morphological disturbance. For the three last ones, our interpretation is that they decrease the time of distant interaction between the two galaxies before the merger occurs. A smaller impact parameter or velocity obviously leads to a faster merger. A small inclination of the orbital plane can have the same effect, for it triggers tidal effects (as shown by the formation of long tails), which help to remove angular momentum. Indeed, one can see in Table~\ref{runs} that the smaller the merging time, the larger the disk thickening and bulge masses. The distant interaction before the merger mainly disturbs the smallest galaxy. When the merging time is large, the smaller galaxy is more dispersed before the merger occurs, thus its effects on the main galaxy during the merger itself are smaller. For instance, we have checked that the companion is significantly more dispersed on prograde orbits than on retrograde orbits. For the 4.5:1 mergers, at the moment when the companion enters the stellar disk radius of the main galaxy, the companion mass still included in its initial radius is smaller by 34\% on average when the orbit is prograde (depending on other parameters). Therefore, it induces less thickening and fuels the bulge component less efficiently.

That a retrograde orbit disturbs the main galaxy more than a prograde one may seem surprising, since a prograde orbit induces larger tidal perturbations\footnote{We have only studied orbits than are prograde for both galaxies, or retrograde for both galaxies. This result could be different if the orbit is prograde for one of the galaxies and retrograde for the other one.}. A visual inspection of some simulations leads us to the following interpretation: on a prograde orbit, the companion is rapidly dispersed by the tidal forces ; it exerts tidal forces at large distances at the beginning of the interaction, but is later on too dispersed to strongly disturb the main galaxy at short distances. On retrograde orbits, the companion is more compact when it gets close to the main galaxy, because it has undergone smaller tidal effects, it can then induce stronger perturbations on the merger remnant. To confirm this interpretation, we have defined a ''tidal parameter'' $T$ to describe the effects of the interaction on each galaxy. We could first define it as:
\begin{equation}
T=\frac{F_{\mathrm{tidal}}}{F_{\mathrm{int}}}
\times
\frac{1}{\left| \omega_{\mathrm{orb}} - \omega_{\mathrm{int}} \right|}
\end{equation}

For each galaxy, the tidal and internal forces $F_{\mathrm{tidal}}$ and $F_{\mathrm{int}}$ are mean values over the outer stellar disk, $\omega_{\mathrm{int}}$ is the rotation frequency measured at this radius, and $\omega_{\mathrm{orb}}$ is the orbital pattern speed of the other galaxy. The second factor in this equation represents the time during which a given region of the disk will undergo the same tidal force. For an exact resonance, $\omega_{\mathrm{orb}}=\omega_{\mathrm{int}}$, this time is infinite. This factor is large close to the resonance, and small far from the resonance, which indicates how large the effects of the tidal force on the disk will be. However, there is a saturation of the tidal forces, because they cannot act faster than one dynamical time, which is about $1/\omega_{\mathrm{int}}$. This saturation reduces the effects of the resonance, which leads us to replace the above equation by:
\begin{equation}
T=\frac{F_{\mathrm{tidal}}}{F_{\mathrm{int}}}
\times
\frac{1}{\left| \omega_{\mathrm{orb}} - \omega_{\mathrm{int}} \right| + \omega_{\mathrm{int}}}
\end{equation}
To produce a dimensionless parameter, we express it in units of $1/\omega_{\mathrm{int}}$, which means that we finally rewrite it as:
\begin{equation}
T=\frac{F_{\mathrm{tidal}}}{F_{\mathrm{int}}}
\times
\frac{ \omega_{\mathrm{int}} }{\left| \omega_{\mathrm{orb}} - \omega_{\mathrm{int}} \right| + \omega_{\mathrm{int}}}
\end{equation}

\begin{figure}
\centering
\resizebox{8cm}{!}{\includegraphics{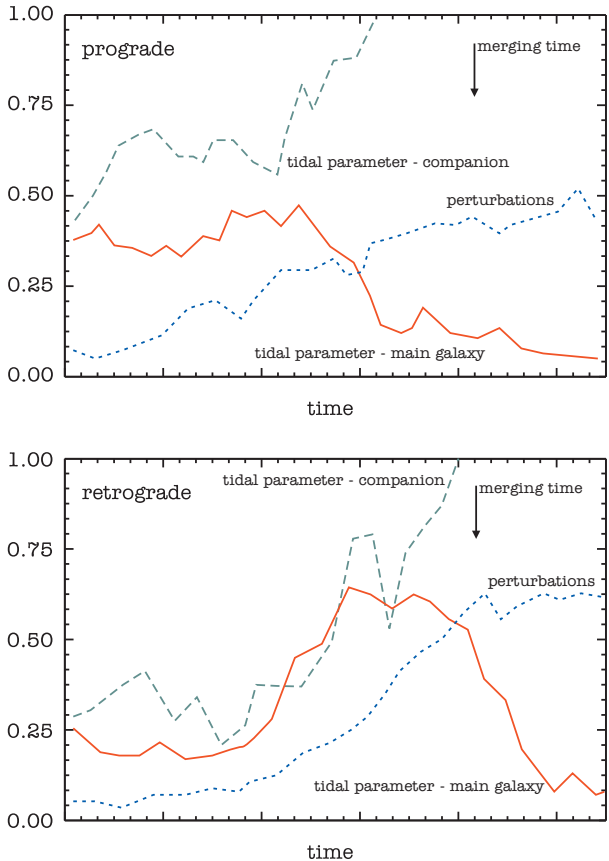}}
\caption{Evolution of the tidal parameter $T$ (see text) for each galaxy and the perturbations of the main galaxy. Red/solid line: tidal parameter of the main galaxy, amplified by the square of the mass ratio. Green/dashed line: tidal parameter of the companion. One can consider that the companion is dispersed when this parameter becomes larger than 1. Blue/dotted line: perturbations of the main galaxy, estimated through the mean value of $\Delta R / R$ (see text). The time axis have been rescaled to allow a comparison of the two figures: the time axis begins when the distance between the two galaxies is 75 kpc, and the arrow indicates the ''merging time'' defined in Sect.~\ref{defs}. The unit time is thus reduced by factor of 0.81 for the retrograde case with respect to the prograde one (the prograde merger is faster).}
\label{tidal_1}
\end{figure}

We show in Fig.~\ref{tidal_1} the results for two 7:1 mergers, with prograde/retrograde orbits and other parameters unchanged (Runs~15 and 16). We also quantify the perturbations induced in the main galaxy. If $R$ is the mean radius (over one rotation) of a star, we have measured its relative variation $\Delta R / R$ averaged over the stellar disk. In Fig.~\ref{tidal_1}, we can consider that the companion has been dispersed when its tidal parameter $T$ becomes larger than 1; the tidal effects are then larger than then internal gravity.

In the prograde case, the companion undergoes strong tidal effects at the beginning of the encounter. It is dispersed rather rapidly, while still located at more than two radii away from the main galaxy. After that, the tidal effects on the main galaxy become much weaker since the companion has been dispersed. The perturbations on the main galaxy are thus mainly initiated by tidal forces exerted at large distances, which leads to a net moderate increase in $\Delta R / R$.

In the retrograde case, the tidal parameters for both galaxies are in the first place about twice smaller. Then, the companion dispersion occurs later, when it is at about 1.5 radii from the main galaxy center. Hence, the companion reaches the main galaxy disk before being dissolved. Then, it causes a strong perturbation on the main galaxy, which is thus affected more strongly than in the direct case. While the increase in $\Delta R / R$ was smaller than in the prograde case during the distant interaction at the beginning of the encounter, here it becomes much larger at the end of the merger, when the companion reaches small radii. The final value of $\Delta R / R$ is 35\% larger than in the prograde case. 

This explains why the main galaxy tends to be more disturbed (regarding its bulge mass and thickening) when the orbit is retrograde. Thus, on a prograde orbit, the companion is largely dispersed by the tidal interaction before colliding with the disk of the galaxy. On a retrograde orbit, it is less dispersed, and hence a significant collision occurs between the disk of the main galaxy and the companion. However, the differences between prograde and retrograde orbits, as well as the variations with other orbital parameters, remain generally smaller than the differences from one mass ratio to another.

\paragraph{Kinematics}
For a given direction of the orbit (prograde/retrograde), the merger remnant has larger velocity dispersions for small impact parameters, small encounter velocities, or small inclinations of the orbital plane. As explained above for the morphological aspect, such parameters lead to a merger with a less-dispersed companion, which induces greater disturbances.

The retrograde orbits lead to systems with larger velocity dispersions than the prograde ones, which can be explained both by the stronger general perturbations described above, and the presence of counter-rotating stars from the companion.

\subsection{Bulge properties}

The bulges of the merger remnants in the 4.5:1--10:1 range of mass ratios have large masses, even if they do not exceed the disk mass. The typical bulge-to-visible mass ratios are 0.20--0.25 for the 10:1 mergers, 0.30--0.40 for the 7:1 ones, and 0.35--0.45 for the 4.5:1 ones (the exact values are mentioned in Table~\ref{runs}). These bulge-to-total mass ratios correspond to bulge-to-disk ratios ranging from 0.35 to 0.8.

The mean ratio of the bulge extent to the disk extent, as defined in L\"utticke et al. (2004), is 0.42$\pm$0.06 for the 10:1 mergers, 0.49$\pm$0.07 for the 7:1 ones, and 0.58$\pm$0.1 for the 4.5:1 ones. In these merger remnants, the bulge represents a large part of the system, in terms of mass as well as in terms of size, especially for the 4.5:1 and 7:1 mass ratios.

When the system is observed close to edge-on, some bulges have disky isophotes, but others have boxy isophotes. The latter is the case for the 7:1 merger shown in Fig.~\ref{maps}. In this system, the bulge extent is 7 kpc, and we measure $<a_4>=-0.018$ between radii 3.5 and 5.5 kpc (see also Fig.~\ref{a4r} for the whole curve of $a_4(r)$ in the bulge and disk components). Such boxy bulges, with a large radial extent and a large mass, could correspond to the ''thick boxy bulges'' reported recently by L\"utticke et al. (2004). Over the whole sample of 4.5:1 and 7:1 mergers, using several edge-on lines-of-sight for each system, we found that 27\% of the bulges appear significantly boxy and 18\% are significantly disky (but most bulges cannot be classified in a robust way because of the limited resolution of our simulations).

\subsection{Gas content}\label{gas}
Even if a significant star-forming event occurs in the galaxy center, and if some gas is removed in tidal tails, remnants of the 4.5--10:1 mergers contain several percent of gas in their disk. In our sample, prograde orbits can lead to the consumption of up to 35\% of the gas of the main galaxy in a central starburst, and can remove up to 55\% of the gas mass in tidal tails. On the other hand, retrograde orbits leave the initial gaseous disk less affected. Moreover, a large fraction of the tidally removed gas falls back on the galaxy, and the companion contributes some gas, too. Due to the dissipational nature of gas, its evolution is different to that of the stars, and unlike stars is less disturbed in retrograde orbits.

Even if some hybrid merger remnants contain less than 2\% of gas, the mean gas fraction in the stellar disk is 3 \% for the 4.5:1 remnants, 3.5 \% for the 7:1 ones and 5 \% for the 10:1 ones. The 4.5:1 and 7:1 remnants thus contain about half the gas of the main parent galaxy, and of the isolated galaxy in the control run: the initial gas mass fraction in the main galaxy is 8\%, and in the control run we find a gas mass fraction of 6.5--7 \% at the time where the merger remnants are analyzed. Thus, the hybrid merger remnants in the 4.5--10:1 range are really gas poorer than the isolated spiral galaxies, but are more gas-rich than the normal ellipticals.

The gas brought in by the companion and returning from tidal tails is generally found at large radii, where it often forms rings. In our set of simulations we have found two polar or strongly inclined rings, and several equatorial rings that will appear as ''dust lanes'' when the system is seen edge-on.

\subsection{Other morphological properties}

About one third of the disky merger remnants have stellar bars with bar strength $Q_b$ up to 0.40. Most of the other ones have oval distortions or lenses. 

The gaseous disk or dust-lanes, that have been strongly disturbed, are generally warped, and sometimes this warp is also visible in the stellar component, even after a few dynamical times. This is for instance the case for the 7:1 merger remnant shown in Fig.~\ref{maps}.

\medskip

Here we have described the main properties of the merger remnants for the intermediate mass ratios. In the next section, we will compare these to the observed  S0 galaxies.


\section{Discussion and implications}

\subsection{Sensitivity of the results on gas physics and star formation}
An important concern regarding simulations of galaxy mergers is whether results are sensitive to the gas dynamical scheme and star formation models, for both are questionable (they do not reproduce exactly the real phenomena occurring in the ISM). The results can only be regarded as robust if they are not affected by the gas cooling and star formation parameters. 

We have repeated one simulation (Run~19) with various values of the elasticity parameter $\beta$ in cloud-cloud collisions, and the exponent $b$ of the generalized Schmidt law for star formation (that assumes that the star formation rate is proportional to the two-dimensional gas density to the exponent $b$). The results are given in Table~\ref{gassf}. As one can see, there are some variations in the large-scale morphological and kinematical properties, but they are small (compared to the whole sample of values that we found when exploring the parameter space, see Table~\ref{runs}), and it seems that these are random variations rather than a systematic dependence of the result on one parameter. 

\begin{table}
\centering
\begin{tabular}{cccccc}
\hline
\hline
$\beta$ & $b$ & & $E$  & $B/T$ & $v/\sigma_{\|}$  \\
\hline
0.8     &  1.4             & & 5.6  & 0.41  & 1.05    \\
0.8     &  1.0             & & 5.5  & 0.42  & 1.01    \\
0.8     &  2.0             & & 5.5  & 0.43  & 1.07    \\
0.6     &  1.4             & & 5.4  & 0.41  & 1.05    \\
0.9     &  1.4             & & 5.6  & 0.39  & 1.03    \\
\hline
\end{tabular}
\caption{Tests of the sensitivity of the results on the gas dynamics and star formation schemes. The physical parameters are that of Run~19, and we vary the elasticity factor of cloud-cloud collisions $\beta=\beta_t=\beta_r$, and the exponent $b$ of the generalized Schmidt law used to computed the star formation rate. The values of the main morphological and kinematical indicators, defined as in Table~\ref{runs} and the rest of the paper, are given.}\label{gassf}
\end{table}

The central regions (less than 500kpc in radius) are much more affected by these parameters. The central density peak can change by a factor of more than two when we vary $b$ from 1 to 2 or $\beta$ from 0.8 to 0.6. Results regarding the central gas infall and central starburst would then be very sensitive to the modeling of the ISM and star formation. However, we have mainly studied large-scale properties of the merger remnants, outside of the inner regions, so we can consider our results as rather robust, without studying in more detail how they are affected by the schemes for gas dynamics and star formation.

\subsection{A formation mechanism for S0 galaxies}\label{so}

Unequal-mass galaxy mergers have been proposed by Bekki (1998) as a mechanism for the formation of S0 galaxies with outer exponential disks. In our simulations, the 4.5:1 and 7:1 merger remnants, and some of the 10:1 remnants, are good candidates for S0 galaxies. Up to now, we had called them ''hybrid'' systems, since they correspond to the ''spiral-like morphology but elliptical-like kinematics'' merger remnants observed in unrelaxed systems by Jog \& Chitre (2002). They are abnormally hot and thick disk galaxies, so they are similar to the S0s. Several properties of the simulated merger remnants can be compared to the observed properties of the S0s in detail as follows:

\begin{itemize}
\item They have a massive and extended disk component, that is not just a faint outer disk.
\item The bulge generally contains 30 to 40\% of the visible mass (bulge-to-total mass ratio), which is compatible with the bulge masses in S0 galaxies (Binney \& Merrifield).
\item They are twice as thick as spiral galaxies: $E\simeq 5.5-7.5$ for intermediate merger remnants instead of $E\simeq8.5$ for the spiral in the control run (see Table~\ref{runs}).
\item They are kinematically hot, with $v/\sigma \simeq 1$ for 4.5:1 mergers, and $v/\sigma \simeq 2$ for 10:1 mergers. Rotation velocities of the order of one to two times the velocity dispersion are in good agreement with the kinematics of observed S0s (e.g., Seifert \& Scorza (1996), Falcon-Barroso et al. (2004), Genzel et al. (2001) for ULIRGS and S0s).
\item Their bulges can be boxy. It even seems that boxy bulges are more frequent than disky ones, but our sample could be too limited to consider this as a definitive result: in most cases, the resolution is too limited to obtain a fair estimation of the bulge diskiness, so that only the most boxy or disky ones can be detected. However, this would be in agreement with the large fraction of boxy bulges observed in S0 galaxies (Seifert \& Scorza 1996, de Souza \& dos Anjos 1987).
\item They are often barred. One third of these merger remnants are found to have strong bars, with bar strength $Q_b \simeq 0.3-0.4$, at the most 0.45. These bar strengths are lower than in spiral galaxies (see e.g., Block et al. 2002 for typical values of $Q_b$ in spiral galaxies): this corresponds to the dilution of the bar gravity torques by the massive bulge, it does not mean that the bars themselves are weak. In most other merger remnants, we find  weak bars, lenses, or oval distortions, with $Q_b \simeq 0.1-0.2$. At the most one fourth of the merger remnants do not have at least a weak bar. S0 galaxies are often barred, too. That bars are present and often strong, while spiral arms are weak or even absent, is a common point between 
the hybrid merger remnants in our simulations and the observed S0 galaxies.
\end{itemize}

Many properties of the S0 galaxies are thus reproduced by the ''hybrid'' merger remnants in the intermediate range of mass ratios. One the other hand, we found two properties that are not well reproduced, but this can be explained:
\begin{itemize}
\item Many S0 galaxies are gas depleted. Our merger remnants contain less gas than before the merger, but most of them still contain a few percent of gas. Yet, the environment could well explain this difference: in our simulations, the merging system is fully isolated. In the reality, many S0s are in clusters or rich environments, so they could have been stripped of their gas by environmental effects. Since we are studying relaxed systems, the merger has occurred at least several dynamical times ago, so that there has been enough time to deplete the gas from these galaxies.
\item S0s have radial scalelengths smaller than spiral galaxies (Binney \& Tremaine 1987). In our simulations, we do not find any systematic difference between the radial scalelength of the pre-existing spiral galaxy and that of the merger remnant. However, we have assumed a constant mass-to-light ratio. Because of the central starburst induced by the merger, we may then under-estimate the central luminosity. A smaller central mass-to-light ratio may decrease the observed radial scalelength, which would be missed in our simulations, and could explain why we failed to reproduce the small radial scalelengths of the S0s. \end{itemize}

Then, even if some properties of the S0s are not well reproduced by our models -- perhaps because of the physical limitation of the models themselves -- the remnants of mergers with intermediate mass ratios look very similar to the S0s. Moreover, these unequal-mass mergers are expected to be frequent, especially at high redshifts. The remnants from these
should then be commonly observed at low redshifts, which is an additional reason to believe that the merger remnants in the range of mass ratios 4.5:1--10:1 are the progenitors of S0 galaxies.

However, there are more S0s observed in clusters of galaxies at $z=0$ than in clusters at $z=1$. This implies that many S0s are formed inside clusters (and not before entering the cluster), while the relative velocity of galaxies in cluster are too high to allow mergers to occur. A first interpretation is that unequal-mass mergers are not the only scenario for the formation of S0 galaxies, but that another independent mechanism forms S0s in clusters, most likely through galaxy harassment (Moore et al. 1996, 1998). Yet, there are also S0 galaxies found outside of clusters, that would still be formed by unequal-mass mergers. Another interpretation is that S0s are the result of unequal-mass mergers, but that the merger is often not enough to form an S0, and additional harassment inside clusters is required to form a real S0 (for instance because there is still gas in the system after the merger, as noticed above). Thus, S0s observed in clusters would be the result of unequal-mass mergers, before they entered clusters, and environmental effects later on, inside the clusters. Probably both interpretations correspond to situations that do occur; the common conclusion is that unequal mass mergers alone cannot have formed all the S0 galaxies. But some S0 galaxies, present in the field or in young clusters, cannot be the result of environmental effects in clusters, and are more probably remnants of unequal-mass mergers.

\subsection{Long-term evolution of merger remnants}
 
\subsubsection{Evolution at high redshifts}
In this paper, and in our earlier work (Bournaud et al. 2004), we have shown that the new mass range 4:1--10:1 reproduces the observed, mixed properties of some peculiar galaxies well (Chitre \& Jog 2002). Since this mass-range is likely to be more common than the equal-mass mergers, especially at high redshifts as shown in the hierarchical merging models (e.g., Steinmetz \& Navarro 2002), we expect that a large fraction of galaxies at high redshifts should be such peculiar systems. This prediction is in agreement with observations of galaxies that show that the galaxy morphology evolves with redshift (Abraham \& van den Bergh 2001).

The galaxy mergers at high redshift may however behave in a different way than in our sample, for galaxies at high redshift contain more gas. This is likely:
\begin{itemize}
\item to reduce the effects the disk destruction or thickening, because of gas falling back after the merger
\item to increase the star formation burst induced both by major mergers and mergers in mass range 4:1-10:1. Indeed, we noticed in Sect.~\ref{gas} than even the 7:1 or 10:1 mergers lead to a noticeable gas consumption. Thus, if the colliding galaxies are gas-rich, this can lead to a major starburst.
\end{itemize}
New simulations with gas-richer galaxies would be required to study the details of the
cosmological importance of the 4:1--10:1 mergers, in particular for the high-redshift starburst.

\subsubsection{Successive mergers}
Since galaxy mergers in the 4.5:1--10:1 range are expected to be common, it is likely that some systems have undergone several mergers of this kind. It is even more likely for a given galaxy to undergo several unequal-mass mergers than one 1:1 merger. So far we have mainly discussed the outcome of a single merger. Subsequent, multiple unequal-mass mergers could give rise to an 
elliptical remnant.
 This is a different pathway for the formation of an elliptical galaxy compared to the standard, major galaxy merger scenario. We give one example here to illustrate this, but the detailed study of this process is beyond the scope of this paper.

\begin{figure}
\centering
\resizebox{8cm}{!}{\includegraphics{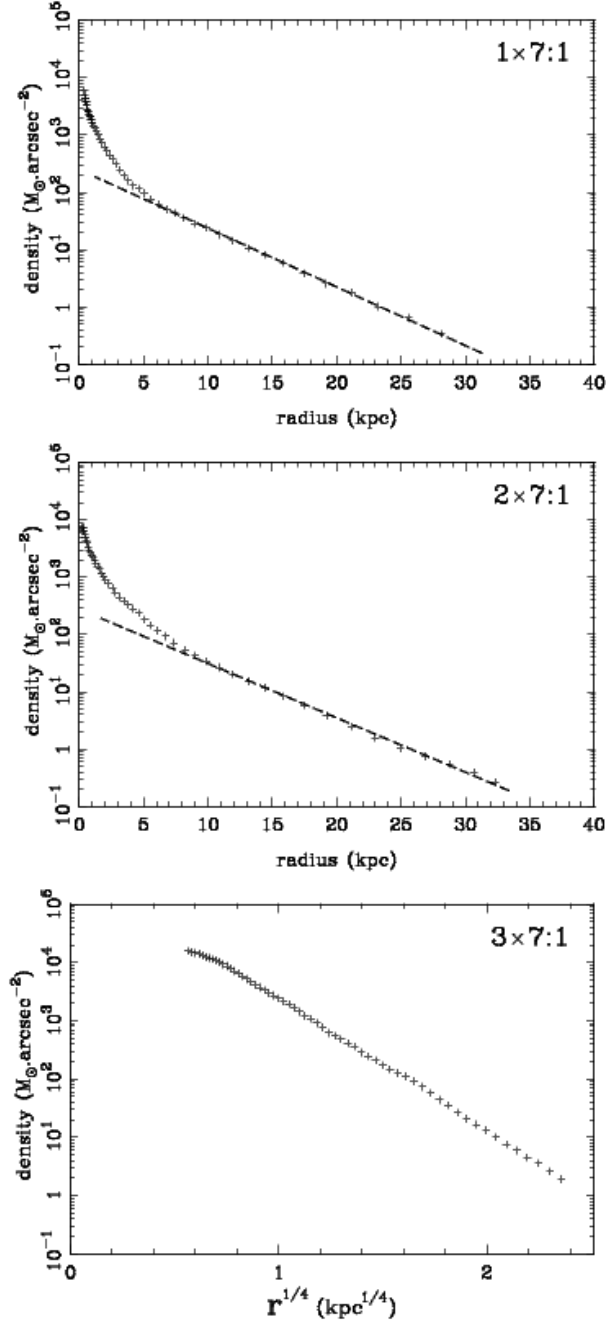}}
\caption{Successive 7:1 mergers: luminosity profiles of the relaxed remnant after one merger (disk galaxy), two mergers (disk galaxy) and three mergers (Elliptical-like morphology). 3.5 Gyrs separate the consecutive mergers.}
\label{multi}
\end{figure}

In Fig.~\ref{multi} we show the result of three successive mergers with mass ratios 7:1. The parameters are $i=33$ degrees, $r=35$ kpc, $V=50$ km~s$^{-1}$. The first and third companions are on prograde orbits, the second one is on a retrograde orbit. Several dynamical times separate each merger. The radial luminosity profiles are shown in Fig.~\ref{multi}: the first 7:1 merger has already been studied. After the second merger, we still observe a robust exponential disk, that is more similar to a 4.5:1 remnant than to a 7:1 one: its flattening is E=5.9, its bulge-to-visible mass ratio 0.36, and its kinematics corresponds to $v/\sigma = 1.1$. After the third merger, no robust exponential disk can be fitted to the luminosity profile any longer, instead
 the mass distribution can be well fitted by an $r^{1/4}$ profile. This  remnant of the multiple mergers is an E5 elliptical galaxy when observed under the projection that gives the largest flattening, with disky isophotes ($a_4 = 0.016$) and $v/\sigma = 0.70$. This example shows that several subsequent mergers in the mass ratio 4.5:1--10:1 can lead to the formation of an elliptical-like object. We have run several other examples where an elliptical-like object is formed by two 4.5:1 mergers or three 7:1 mergers. The detailed analysis of these simulations, and the comparison with major mergers remnants and observed elliptical galaxies, will be the subject of a forthcoming paper. Yet, it is important to notice that this multiple-merger mechanism for the formation of elliptical galaxies can be more frequent than the scenario of a single, major merger: we have estimated this using the GalICS/MoMaF database of galaxies{\footnote {\tt http://galics.iap.fr}}. We have selected 1000 galaxies with stellar masses higher than $4\times 10^{10}$~M$_{\sun}$, and followed their merger history from $z=0$ to $z=0.6$. We find that for these galaxies and in this redshift range, mergers in the 4:1--10:1 range of mass ratios are 6.5 times more frequent than major mergers in the 1:1--3:1 range. This can vary with redshift and with the mass of galaxies, but three successive intermediate (4:1--10:1) mergers are as likely
as or even more likely than one single major (1:1--3:1) merger.

This new scenario may explain the formation of the giant boxy elliptical galaxies, which the single major merger scenario cannot account for (Naab \& Burkert 2003).

\subsubsection{Gas accretion}
It is also possible that a remnant of an unequal-mass merger further accretes a large amount of gas and thus forms a thin, kinematically cold spiral disk in a few Gyrs (Block et al. 2002). These could then evolve into a normal spiral galaxy embedded in a
thick and kinematically hot disk. Indeed, the merger remnants in the 4:1--10:1 range, that are 2 or even 3 times thicker than spiral galaxies, with large velocity dispersions, could be the progenitors of the thick disks observed around some spiral galaxies,
 as is the case around the Milky Way. However a detailed study of this scenario, and the comparison with observed thick disks, remains to be performed.

\subsection{Discussion of the classification criterion}

According to the criterion detailed above, we have classified as ''disk'' a galaxy that has an exponential profile, and as ''elliptical'' a galaxy that is best fitted by a de Vaucouleurs profile. We made this choice to reproduce the most frequently used observational procedure, so that a comparison can be made with observational classifications. However, some mistakes may thereby have been introduced, both in our numerical work and in observational classifications. Indeed, some systems that have an exponential profile may not actually be disks, while some disks with a de Vaucouleurs profile may in principle exist. 

We have shown that galaxies classified as ''disk'' on the basis of their exponential radial profile actually have the underlying morphology of a disk. They have a high flattening, higher than in elliptical galaxies (see values of $E$ in Table~2), and their isophotes are highly disky in edge-on projections (see the values of $a_4$ in Fig.~8, and the edge-on projections in Fig.~7). So, from a morphological point of view, they are really disk galaxies. However, depending on the mass ratio, their kinematics is not always typical of spiral galaxies: when the mass ratio is 7:1 or 4.5:1, they can have $v/\sigma$ as small as 1. These systems finally have a disk-like morphology but not spiral-like kinematics: they have been described in Bournaud et al. (2004) and we called them ''hybrid'' systems. Even if their global kinematics is very hot, their mean rotation axis remains aligned with the morphological disk axis. For instance in our 7:1 merger remnants, we measured the mean angle between the rotation axis and the morphological flattening axis smaller than 10 degrees: their kinematical properties are not completely independant of their disk-like morphology. Thus, in our sample of massive merger remnants, the criterion based on the exponential profile selects galaxies that are actually disk-like galaxies (but this does not imply that they also have spiral-like kinematics).

Reciprocally, a system with a de Vaucouleurs profile (then classified as ''elliptical'' in our sample) may in principle have a disky morphology, rather than being an elliptical-like spheroid. However, all of the systems showing a de Vaucouleurs profile are not as flat as disks (see Table~2). Their isophotes are sometimes disky but the values of $a_4$ (see Sect.~4 and Fig.~8) are typical of the observed ''disky ellipticals'': they are much less disky than real disk galaxies. Also, these systems have $v/\sigma$ close to 1 or smaller (Table~2). Thus, all the systems with a de Vaucouleurs profile in our sample have both the morphological and kinematical properties of elliptical galaxies: no stellar disks with a de Vaucouleur profile is formed in the merger of massive spiral galaxies.

Thus, the classification criterion based on the radial luminosity profiles seems to provide a fair indicator of whether a galaxy is a disk galaxy or an elliptical galaxy.

\subsection{Comparison with other works}

In this paper, we have explored the parameter space in detail, especially for the new range of mass-ratios, and obtained the main morphological and kinematics properties of the remnants. Although a transition from an elliptical to a disk-like behavior in the remnants as one goes from 1:1 to 10:1 was expected based on previous works in the literature, it was not expected that the remnants for the range 4:1-10:1 would have hybrid behavior. Certainly, the fact that the mergers show hot kinematics already at 10:1 or 7:1 but show an elliptical-like mass profile around 4:1 is a new and a surprising result from our work.

The mergers for the mass range 3:1-4:1 have been studied in the past (e.g., Naab, Burkert \& Hernquist 1999, Barnes 1998, Bendo \& Barnes 2000, Naab \& Burkert 2003). However, these papers do not explicitly consider the radial mass profiles but instead consider the diskiness of the projection of the remnant, where the diskiness is denoted by $a_4$ or the coefficient of the $\cos 4 \phi $ term in the Fourier expansion (defined in Section 2.3).
On the other hand, we have studied a proper radial mass distribution in this paper. For some cases in this range, Naab \& Burkert (2003) do find a disky behavior of the remnant. However, it is not clear that there is a one-to-one correspondence between diskiness as defined by $a_4 > 0$ and a disk distribution as defined by an exponential surface density distribution as observed in isolated spirals. For example, in the study of 27 advanced mergers, Chitre \& Jog (2002) found that some galaxies with an
 outer exponential disk distribution showed boxiness ($a_4 < 0$) as in AM 2146-350, and vice versa when a merger which showed a clear $r^{1/4}$ elliptical-like fit gave a disky value as seen from $a_4$ as in Arp 193. Moreover, even within a galaxy, the remnant can change from diskiness to boxiness as one goes from inner to outer region as in Arp 221 or vice versa as in AM038-230 (Chitre \& Jog 2002- see Appendix A), which is also the case in our Run~11 (see Fig.2) and in several other runs. Also, it has been shown that the same merger remnant can appear disky or boxy when viewed from different orientations (Hernquist 1993). Thus there is evidence that $a_4$ is not a completely reliable indicator of true disk behavior. 

Recently, Gonz\'alez-Garc\'ia \& Balcells (2005) have found that, for mass ratios around 3:1, the merger remnant is sometimes an elliptical galaxy and sometimes a disk galaxy. Then, the transition between major mergers forming elliptical galaxies and other mergers resulting in disk galaxies should be around 3:1, which is in agreement with our work that sets this limit between 3:1 and 4.5:1. This also confirms our result that for higher mass ratios like 5:1 or 7:1, even if the merger is not really ''minor'', the stellar disk is not completely destroyed.


\section{Conclusion}

We have explored a new range of mass ratio (4:1--10:1) of galaxy mergers via N-body simulations, and have covered the parameter space extensively for these ratios, which makes
our results statistically significant.
 We have shown that the transition between elliptical and disk-like remnants, as classified both from their radial profiles and their vertical mass distribution, occurs for a well-defined range of mass ratios, between 3:1 and 4.5:1. Yet, the mergers in the range 4:1--10:1 do not result in disturbed spiral galaxies, but instead they result
 in hybrid remnants that have the morphology of a disk galaxy with very hot, or even elliptical-like, kinematics, as seen in our
preliminary study (Bournaud et al. 2004). These peculiar systems seem to reproduce well
the observed properties of the systems analyzed
 by Jog \& Chitre (2002). These remnants can be considered as good candidates for S0 galaxies for they reproduce most of the S0 properties. However, as discussed
  at the end of Sect.~\ref{so}, this cannot explain the formation of all the S0s (at least in clusters), and other mechanisms must play a role in the formation of S0s, either after unequal mass mergers have occurred, or as alternative
formation mechanisms that do not require any merger. The study of the orbits and the details of relaxation, especially for the transition region between disk-like and elliptical remnants for mass ratios around 4:1, will be pursued in a future paper. 

We have also studied the influence of orbital parameters on the merger, but found that the most important parameter is the mass ratio. We then define three classes of galaxy mergers: the {\it major mergers} (1:1--4:1) that form elliptical galaxies, the {\it intermediate mergers} (4:1--10:1) that form peculiar remnants that could be the progenitors of S0 galaxies, and the {\it minor mergers} (more than 10:1) that result in disturbed spiral galaxies. The mass ratios 
quoted here are the ratios of the stellar masses.

Since they are expected to be very frequent, especially at high redshifts, the intermediate mergers may explain not only the formation of S0 galaxies, but also of elliptical galaxies after several subsequent intermediate mergers, instead of one single major merger, and of thick disks surrounding younger, cold, spiral disks.

\begin{acknowledgements}
We acknowledge the anonymous referee for valuable comments. The N-body simulations in this work were computed on the Fujitsu NEC-SX5 of the CNRS computing center, at IDRIS. This work uses the GalICS/MoMaF Database of Galaxies ({\tt http://galics.iap.fr}). We are happy to acknowledge the support of the Indo-French grant IFCPAR/2704-1.
\end{acknowledgements}


\end{document}